\documentclass[aps,prd,showpacs,floats,letterpaper,floatfix,
groupedaddress,superscriptaddress
,nofootinbib
,eqsecnum
,twocolumn
]{revtex4}
\usepackage{amsmath}
\usepackage{amsfonts}
\usepackage{bm}
\usepackage{color}
\usepackage{graphicx}
  \newcommand{\apjl}{Astrophys. J. Lett.}

  \newcommand{\lrr}{Living Rev. Relativ.}

\def\v1v2{{\bf v}_1 \cdot {\bf v}_2}

\allowdisplaybreaks

\begin{document}

\title{Equations of motion for compact binary systems in general relativity: Do they depend on the bodies' internal structure at the third post-Newtonian order?}

\author{
Clifford M.~Will} \email{cmw@phys.ufl.edu}
\affiliation{Department of Physics, University of Florida, Gainesville, Florida 32611, USA}
\affiliation{GReCO, Institut d'Astrophysique de Paris, CNRS,\\ 
Sorbonne Universit\'e, 98 bis Boulevard Arago, 75014 Paris, France}

\date{\today}

\begin{abstract}
We present and discuss the possibility, derived from work carried out 20 years ago, that the equations of motion for compact binary neutron stars at the third post-Newtonian (3PN) order in general relativity might actually depend on the internal structure of the bodies.  These effects involve integrals over the density and internal gravitational potentials of the bodies that are independent of the mass and radius of the bodies, but dependent on their equations of state.  These effects could alter the coefficients in the 3PN equations derived using ``point mass'' methods by as much as 100 percent.   They were found in independent calculations done at Washington University using the Direct Integration of the Relaxed Einstein Equations (DIRE) approach, and at the Institut d'Astrophysique de Paris using the Multipolar post-Minkowskian (MPPM) approach.  Neither calculation was completed because of the enormous complexity of the algebraic computations and the limitations of software of the day (Maple, Mathematica), and because of an assumption that the effects would somehow cancel or be removable by some transformation.  This assumption was rooted in the Strong Equivalence Principle (SEP), which would suppress such effects up to the stage where tidal interactions become important, effectively 5PN order for compact bodies.  SEP was well supported at lower PN orders and in special examples.  We argue that this assumption needs to be verified by actual calculations. If the results show that these structure-dependent terms exactly cancel or can be absorbed into renormalized masses or shifted positions of each body, this would provide remarkable support for the Strong Equivalence Principle of general relativity.  But if they do not cancel and are not incorporated into gravitational waveforms, they could severely impact efforts using next-generation gravitational-wave interferometers to extract information about the equation of state for neutron star matter from gravitational-wave signals from binary neutron star or black hole-neutron star mergers.  
\end{abstract}

\maketitle

\section{Introduction and summary}
\label{sec:intro}

Gravitational-wave science has reached a level of maturity and excitement that have exceeded even the most optimistic expectations of gravitational physicists at the turn of the millennium.  The network of laser interferometric gravitational observatories, LIGO, Virgo and Kagra, have detected more than 80 binary black hole mergers, two double neutron star mergers and three black-hole neutron-star mergers \cite{2023PhRvX..13d1039A}.  Pulsar timing array collaborations have reported evidence for a stochastic background of ``nanohertz'' gravitational waves \cite{2023ApJ...951L...8A}.    
The space interferometer LISA is on track for construction and launch in the mid-2030's.
Planning has begun for design and construction of ``next-generation'' ground-based interferometers, such as Einstein Telescope and Cosmic Explorer, to make observations with roughly 10 times the strain sensitivity of the current advanced devices \cite{2023arXiv230613745E,Maggiore_2020}.

For the ground-based interferometers, the process of detecting gravitational waves and extracting useful science from them relies upon theoretical template waveforms calculated from general relativity that are as accurate and faithful to the ``true'' waveform as can be practicably achieved.  In the early inspiral regime, the waveforms are obtained using post-Newtonian (PN) theory, which is, roughly speaking, an expansion of Einstein's equations in powers of $(v/c)^2 \sim (Gm/rc^2)$, where $m$, $v$ and $r$ are the characteristic mass, velocity and separation within the system and $G$ and $c$ are the gravitational constant and speed of light (for a recent review of these methods, see \cite{2024LRR....27....4B}).  In the strong-field merger regime, numerical relativity must be used, and in the post-merger regime, black hole perturbation theory must be used, to obtain the quasinormal mode ``ringdown'' waveforms.   To connect these regimes, techniques such as the ``effective one-body'' approach have been important.  The success of this approach has been undeniable, with the numerous detections of mergers, the measurement of masses and spins, the estimates of populations and formation channels, and so on.  

The inspiral part of the waveforms contain an assumption that often goes unstated, but that is very important, namely that the equations of motion and the gravitational waveforms of non-spinning bodies are independent of the internal structure of the bodies (for rotating bodies the effects of spin can be included by established techniques).  This assumption is not absolute, of course, because it is well recognized that bodies exert tidal forces on one another, and the response to these forces (as in the Earth-Moon system, for example) depends strongly on the bodies' internal structure.   But black holes and neutron stars are extremely small, and therefore tidal interactions are not expected to be important until very late in the inspiral (and for black holes, probably not at all, at least until the stage when numerical relativity must be invoked).  But when and if these effects are detected in the late part of the inspiral, tidal effects are expected to play the leading role in using gravitational wave signals to study the equation of state of nuclear matter at neutron star densities (for a review, see \cite{2020GReGr..52..109C}; for an early result using data from the binary neutron star merger GW170817 see \cite{2017PhRvL.119p1101A}). The effects arise from standard Newtonian gravity, but in the context of compact binary systems, they are viewed as being of ``fifth'' post-Newtonian, or 5PN order (i.e. of order $(Gm/rc^2)^5$) relative to Newtonian theory.  This is because the effects of a quadrupole moment of a body on the motion of its companion scale as $(s/r)^2$ relative to that of a spherical body, where $s$ is a measure of the body's size, and the size of that quadrupole moment resulting from the tidal effect of the companion scales as  $(s/r)^3$; the combined effect scales as $(s/r)^5$.  But since $s \sim Gm/c^2$ for compact bodies, the {\em effective} size is of 5PN order.

In summary, the conventional assumption has been that for non-spinning compact bodies, the motion and gravitational waves are independent of the bodies' internal structure until effectively 5PN order. Damour proposed an ``effacement'' principle \cite{Damour300}, by which such effects would (like the Cheshire cat) be effaced until that effective order.  Within general relativity, this assumption is not unreasonable, because of a concept known as the Strong Equivalence Principle. 
 
The Weak Principle of Equivalence, the postulate that ``test'' bodies move in a gravitational field in a manner that is independent of their internal structure and composition is a cornerstone of general relativity and all of its modern competitors.  It has become a foundation of the concept of gravity as a manifestation of the curvature of spacetime and has been verified to high precision by numerous experiments \cite{tegp2}, most recently to parts in $10^{15}$ by the space experiment MICROSCOPE \cite{PhysRevLett.119.231101}.

A stronger version of this principle asserts that even gravitationally self-interacting ``test'' bodies, such as planets, stars and black holes, move in the same ``structure independent'' manner (apart from spin effects, of course).  This version, part of the so-called Strong Equivalence Principle, is so strong that general relativity may be the only metric theory of gravity that actually obeys it.  Scalar-tensor theory, its most popular alternative, violates it.  It too has been verified by a wide range of experiments, including lunar laser ranging \cite{tegp2}, and observations of the pulsar J0337+1715 in a triple-star system \cite{2014Natur.505..520R}.   In the post-Newtonian approximation, general relativity's adherence to SEP has been explicitly verified theoretically to 1PN order (within the parametrized post-Newtonian formalism) \cite{1968PhRv..169.1017N,1971ApJ...163..611W} and to 2PN order \cite{2007PhRvD..75l4025M} (see also \cite{1983SvAL....9..230G}).

However, there may be reason to worry that SEP is {\bf not} obeyed at 3PN order, and that internal structure-dependent effects may arise in gravitational waveforms already at 3PN order, two PN orders ``earlier'' than the conventional tidal effects.

The source of this worry is work done in our research group WUGRAV at Washington University in St. Louis during the period 2003 - 04.  We had previously developed a method called DIRE, Direct Integration of the Relaxed Einstein equations, for obtaining equations of motion and gravitational waveforms in a PN expansion \cite{1996PhRvD..54.4813W,2000PhRvD..62l4015P,2002PhRvD..65j4008P}.  A somewhat different approach, called Multipolar post-Minkowskian (MPPM) theory had been pioneered by Blanchet and Damour \cite{1986RSPSA.320..379B,1987RSPSA.409..383B} in Paris (see \cite{2024LRR....27....4B} for a review of all the various methods).  Both methods were based on the same foundation of the Landau-Lifshitz formulation of the Einstein Equations, which leads to a flat-spacetime wave equation for the fields with a source energy-momentum tensor containing both matter and field contributions (see \cite{PW2014} for a pedagogical treatment).  But their approaches to finding solutions were different.  The DIRE approach carried out the integration of the source divided by $|{\bm x} - {\bm x}'|$ over the past flat-spacetime null cone, changing integration variables at the boundary between the near zone and far zone to make the outer integrals explicitly finite, and verifying that the answers were independent of the radius ${\cal R}$ of the near zone.  The MPPM approach solved two problems, a conventional post-Newtonian solution within the near zone (very similar to the DIRE approach), and a multipole-expanded solution of the {\em vacuum} Einstein equations in the far zone, followed by an asymptotic matching procedure to link the two solutions at the near zone-far zone boundary.  The treatments of the interface between the near and far zones in the two methods, while conceptually very different, were shown to be equivalent (see Sec.\ 4.3 of \cite{2014LRR....17....2B}).  These approaches led to the successful calculation by the two groups of the gravitational waveform to 2PN order in 1995, in complete agreement with each other \cite{1995PhRvL..74.3515B} (see also \cite{1995PhRvD..51.5360B,1996PhRvD..54.4813W}).
But in short order, it became clear that 2PN order would not suffice for gravitational-wave detection by the coming advanced ground-based interferometers, and that 3PN order would be needed (and eventually even PN orders beyond that).  

It was here that the two approaches  diverged.  In their early 2PN work, both groups had treated the bodies as point masses (delta functions), and had used various techniques to eliminate or regularize the inevitable infinities that occur in a theory with non-linear field interactions.  In moving to 3PN order, the MPPM team continued with delta function sources, but had to introduce more sophisticated regularization methods, notably dimensional regularization, a method imported from quantum field theory.  In our work using DIRE (with Michael Pati), we elected to treat each body as a finite ball of self-gravitating matter of characteristic size $s$, to expand the various fields about the center of mass of each ball in powers of $s$, and to include the body's ``self'' field, which would scale as $1/s$.  For the equations of motion, we integrated over each body, discarding terms that scaled as negative powers of $s$; these would be the analogues of the ``infinities'' handled by dimensional regularization in the MPPM approach.  We also discarded terms scaling as positive powers of $s$; these would get smaller as the body shrank.  This procedure led to 2PN equations of motion that agreed with earlier results.

As we turned to the 3PN terms using the same procedure, we began to encounter terms involving structure-dependent factors that scaled as $s^0$, that neither grew nor became smaller as the body shrank.  After factoring out suitable powers of the mass of the given body, the factors were dimensionless, independent of either the mass or radius of the body, but dependent on its equation of state.  
  The origin of many these terms was a set of ``triangle'' potentials, that depend on the field point and on {\em two} source points, but they also arose in the simplest of 3PN terms, such as a term proportional to $U^3 {\bm \nabla}U$, where $U$ is the Newtonian potential.   By the time the project had to be ended in 2004, 40 distinct coefficients had been identified (see Table \ref{tab:coeffs} below), and there was no end in sight.  
  
  Ironically, the MPPM group had simultaneously discovered the same problem.  While grappling with the problem of regularizing the singularities in the point-mass approach (before turning to dimensional regularization), Blanchet and Esposito-Far\`ese had worked on the equations of motion for extended bodies, finding complicated terms dependent on the internal structure.  They conjectured that the terms would be eliminated by a change of coordinates, and would help determine the value of a certain ``ambiguity parameter'' that was present in the calculations at that time, and furthermore that this parameter would be independent of the internal structure of the bodies.  But they too, were never able to finalize the project and prove this conjecture.  
  
On the other hand, contemporaneous work by Itoh and Futamase \cite{2003PhRvD..68l1501I,2004PhRvD..69f4018I} seemed to imply that the 3PN equations of motion would be free of structure-dependent effects.  Their approach, called the Strong-Field Point-Particle Limit \cite{2000PhRvD..62f4002I}, surrounded each body by a sphere and carried out surface integrals of field quantities on those spheres.   The  approach was reminiscent of the famous work of Einstein, Infeld and Hoffman (EIH) \cite{1938AnMat..39...65E}.   The idea is to exploit the fact that the metric outside a static spherical body in its momentary rest frame is close to the Schwarzschild metric, which is independent of the body's internal structure, and to use matching conditions and surface integrals to obtain conditions on the motion of the frame that translate into equations of motion.  The composition of the source never enters the calculation.  Itoh and Futamase successfully obtainied the 3PN equations of motion in agreement with other methods.  They even obtained a definite value for the ambiguity parameter that was in agreement with the value that was ultimately fixed in the MPPM method once dimensional regularization was employed.   
However a key part of the Itoh-Futamase method was to make the radius of the sphere surrounding each body be proportional to a small parameter $\epsilon$ and to expand about the limit $\epsilon \to 0$, hence the ``point-particle limit''.

The question is, for a neutron star with a finite radius, does this limiting procedure still faithfully adhere to the original principle that the local geometry outside the star is Schwarzschild, or does it somehow suppress or overlook internal structure effects related to the finite size of the body that might otherwise have been present?   The question may also be applied to other methods that successfully obtained 3PN equations of motion using various regularization techniques, such as the ADM-Hamiltonian approach \cite{jaraschaefer98,jaranowski} and the ``Effective Field Theory'' (EFT) approach
\cite{2011PhRvD..84d4031F}.
  
In this paper, we present the central issues encountered in the DIRE approach, in the hope of reviving interest in the problem of structure dependence in the equations of motion.  Perhaps, armed with modern tools such as machine learning or artificial intelligence, researchers will be able to complete the work and answer the question posed in the title.

In Sec.\ \ref{sec2}, we review the approach use by DIRE for analysing the continuum equations of motion at high PN orders.  Section \ref{sec3} presents two characteristic examples of terms that lead to structure-dependent effect that scale as $s^0$, describes the challenges in working with more complex PN potentials (known as quadrangle potentials), discusses the ``structure coefficients'', and presents typical numerical values for a range of equations of state.  In Sec.\ 
\ref{sec:conclusions} we discuss the results.  In an Appendix we present the complete continuum equations of motion through 3PN order (excluding the well-known 2.5PN radiation-reaction terms) in a form suitable for launching a direct attack on the problem.

We use units in which $G=c=1$, summation over repeated Greek or Latin indices is assumed, and where parentheses (square brackets) around indices denote symmetrization (antisymmetrization).  For spatial indices, there is no distinction between covariant or contravariant placement, since the method adopts a Cartesian metric for spatial slices.  Capital indices denote a product of vector, viz. $\hat{n}^L = \hat{n}^i \hat{n}^j \hat{n}^k \dots$, and the object $\hat{n}^{\langle L \rangle}$ denotes a symmetric trace-free combination (see \cite{PW2014} for definitions and discussion).   Commas in subscripts or superscripts denote partial derivatives and semicolons denote covariant derivatives; dots over functions denote time derivatives.

\section{Foundations for the fluid equations of motion at 3PN order}
\label{sec2}

\subsection{Review of the DIRE method}

DIRE (Direct Integration of the Relaxed Einstein Equations) is a version of the standard post-Minkowskian and post-Newtonian methods developed in \cite{1996PhRvD..54.4813W,2000PhRvD..62l4015P,2002PhRvD..65j4008P}, and summarized in \cite{PW2014}.   One defines the field $h^{\alpha\beta} \equiv \eta^{\alpha\beta} - (-g)^{1/2} g^{\alpha\beta}$, where $\eta^{\alpha\beta}$ is the Minkowski metric and $g$ is the determinant of the spacetime metric $g_{\alpha\beta}$.  In  harmonic coordinates defined by the gauge condition
${h^{\alpha \beta}}_{,\beta} = 0$,  
the Einstein equations take the
form
\begin{eqnarray}
\Box h^{ \alpha \beta } = -16 \pi {\tau}^{ \alpha \beta } \; ,
\label{relaxed}
\end{eqnarray}
where $\Box $
is the flat-spacetime wave operator, and ${\tau}^{ \alpha \beta }$ is made
up of the material stress-energy tensor $T^{ \alpha \beta }$ 
and the contribution of all the
non-linear terms in Einstein's
equations, contained in the Landau-Lifshitz pseudotensor $t_{\rm LL}^{ \alpha \beta }$ and the ``harmonic'' pseudotensor
$t_{\rm H}^{ \alpha \beta }$ (see Eqs.\ (6.5) and (6.53) of  \cite{PW2014} for definitions).  By virtue of the harmonic gauge condition, $\tau^{\alpha\beta}$  satisfies ${\tau^{\alpha \beta}}_{,\beta} = 0$; this is equivalent to the covariant condition ${T^{\alpha \beta}}_{;\beta} = 0$.  

The formal solution of Eq.\ (\ref{relaxed}) with an outgoing wave boundary condition is given by
\begin{equation}
h^{ \alpha \beta }(t,{\bm x}) = 4 \int_{\cal C} \frac{{\tau}^{ \alpha \beta } (t-|{\bm x}-{\bm x}'|,{\bm x}')}{|{\bm x}-{\bm x}'|} d^3x' \,,
\end{equation}
where $\cal C$ is the past flat spacetime null cone whose vertex is the field point $(t,{\bm x})$.
For field points within the near zone, a sphere of radius ${\cal R} \sim \lambda$, where $\lambda$ is the characteristic gravitational wavelength, the time dependence can be expanded, to yield
\begin{align}
h_{\cal N}^{\alpha \beta} (t,{\bm x}) &=  4 \sum_{m=0}^\infty {1
\over {m!}} {\partial^m \over {\partial t^m}} \int_{\cal M}
\tau^{\alpha \beta} (t,{\bm x}^\prime) 
 | {\bm x} - {\bm x^\prime} |^{m-1} d^3 x^\prime \;,
\label{nearexpand}
\end{align}
where $\cal N$ denotes the part of the null cone inside the near-zone world tube, and $\cal M$ denotes the intersection of the hypersurface $t=$
constant with the near-zone world-tube (see Fig.\  (6.3) of \cite{PW2014}).  The integral over the remainder ${\cal C} - {\cal N}$ of the null cone contributes so-called ``tail terms'', and will not be relevant for this discussion (see Sec.\ 6.3.5 of  \cite{PW2014} for discussion of the integration over ${\cal C} - {\cal N}$).

In DIRE I \cite{2000PhRvD..62l4015P}, we modeled the matter sources as balls of pressureless fluid that are spherically symmetric in their comoving inertial reference frames, with an energy-momentum tensor of the form
\begin{equation}
T^{\alpha \beta} = \rho^* \frac{u^\alpha u^\beta}{u^0 \sqrt{-g}} \,,
\label{eq:Tab}
\end{equation}
where $\rho^*$ is the so-called ``conserved'' or baryon density, related to the local mass density $\rho$ by $\rho^* = \rho \sqrt{-g} u^0$; it satisfies the exact continuity equation $\partial \rho^* /\partial t + \bm{\nabla} \cdot (\rho^* {\bm v} ) = 0$, where $v^j = u^j/u^0$.

Following the convention of Blanchet and Damour \cite{1989AIHPA..50..377B}, we defined the ``sigma'' densities,
\begin{equation}
\sigma  \equiv T^{00} + T^{ii} \,, \quad
\sigma^i  \equiv T^{0i} \,,\quad 
\sigma^{ij}  \equiv T^{ij} \,.
\end{equation}
We then obtained formal solutions of Eq.\ (\ref{nearexpand}) in a post-Newtonian expansion, in which $m/r \sim v^2 \sim \epsilon$, $\partial/\partial t \sim v \cdot \nabla$, resulting in $h^{00}$ to order $\epsilon^{9/2}$, $h^{0j}$ to order $\epsilon^{4}$, $h^{jj}$ to order $\epsilon^{9/2}$, and $h^{ij}$ to order $\epsilon^{7/2}$, which included 3.5PN radiation-reaction terms.  These were expressed in terms of instantaneous Poisson-like
potentials and
their generalizations, sometimes called superpotentials.  For a source
$f$, we defined the Poisson potential, superpotential, and
superduperpotential to be 
\begin{align}
P(f) &\equiv {1 \over {4\pi}} \int_{\cal M} {{f(t,{\bm x}^\prime)}
\over {|{\bm x}-{\bm x}^\prime | }} d^3x^\prime \,, \quad \quad \nabla^2
P(f) = -f \,, 
\nonumber \\
S(f)&\equiv {1 \over {4\pi}} \int_{\cal M}f(t,{\bm x}^\prime)|{\bm
x}-{\bm x}^\prime | d^3x^\prime \,,  \nabla^2 S(f) = 2P(f) \,,
\nonumber \\
Sd(f)&\equiv  {1 \over {4\pi}} \int_{\cal M}f(t,{\bm x}^\prime)|{\bm
x}-{\bm x}^\prime |^3 d^3x^\prime ,  \nabla^2 Sd(f) = 12S(f) \,,
\label{definepoisson}
\end{align}
where $f$ could be one of the sigma densities or a complicated function of the various potentials.  At Newtonian order, the solution was 
\begin{equation}
h^{00} = 4U_\sigma = 4\int_{\cal M} \frac{\sigma(t,{\bm x}')}{|{\bm x} - {\bm x}'|} d^3x' \,.
\end{equation}

Explicit formulae for the $h^{\alpha \beta}$ in terms of these sigma-potentials through 3.5PN order can be found in DIRE I, Eqs.\ (5.2), (5.4), (5.8), (5.10), (6.2) and (6.4).
The expansions actually  included 4PN ``tail'' terms from the wave-zone part of the integrals over the past null cone.  These and the radiation reaction terms will not be discussed here. 

The next step in obtaining equations of motion was to relate the sigma densities to $\rho^*$ in a PN expansion, and to iteratively convert all sigma potentials to potentials expressed in terms of $\rho^*$ to the required PN order.   

With the energy-momentum defined as in Eq.\ (\ref{eq:Tab}), the equation of motion for each fluid element is simply the geodesic equation, given in coordinate form by
\begin{equation}
dv^i/dt = -\Gamma^i_{\alpha\beta} v^\alpha v^\beta +
\Gamma^0_{\alpha\beta} v^\alpha v^\beta v^i \,,
\label{geodesic}
\end{equation}
where $\Gamma^\gamma_{\alpha\beta}$ are Christoffel symbols computed
from the metric.  Substituting the solutions for the fields in terms of $\rho^*$ potentials into the standard formulae for the Christoffel symbols resulted in the continuum equations of motion through 3.5PN order,  
\begin{align}
dv^i /dt   &= U^{,i} + a_{PN}^i + a_{2PN}^i + a_{2.5PN}^i 
\nonumber \\
& \qquad + a_{3PN}^i 
+ a_{3.5PN}^i \,,
\label{eq:fluiddvdt}
\end{align}
where 
\begin{equation}
U = \int_{\cal M} \frac{\rho^*(t,{\bm x}')}{|{\bm x} - {\bm x}'|} d^3x'  \,.
\end{equation}

Expressions for  $a_{PN}^i$, $a_{2PN}^i$, $a_{2.5PN}^i$, and  $a_{3.5PN}^i$ are displayed in DIRE II, Eqs.\ (2.24).  In Appendix \ref{app:A}, we redisplay $a_{PN}^i$ and $a_{2PN}^i$, and, for the first time, the significantly more complicated $a_{3PN}^i$.

\subsection{The zoo of post-Newtonian potentials}
\label{sec:zoo}

Newtonian gravity is defined by the potential $U$, which is constructed from the two-point function $1/|{\bm x} - {\bm x}'|$, where $\bm x$ is a field point and ${\bm x}'$ is a source point attached to a mass element.   At 1PN order, all the potentials are based on $\rho^*$ and $|{\bm x} - {\bm x}'|$, of the form
\begin{align}
\Sigma (f) &\equiv \int_{\cal M} {{\rho^*(t,{\bm x}^\prime)f(t,{\bm
x}^\prime)}
\over {|{\bm x}-{\bm x}^\prime | }} d^3x^\prime  \,,
\nonumber \\
X(f)  &\equiv \int_{\cal M} {\rho^*(t,{\bm x}^\prime)f(t,{\bm
x}^\prime)}
{|{\bm x}-{\bm x}^\prime | } d^3x^\prime  \,,
\label{eq:potentials0}
\end{align}
where $X$ is known as a superpotential.  The specific 1PN functions are
\begin{align}
U &\equiv \Sigma(1) \,,  \qquad V^i \equiv \Sigma(v^i) \,, \qquad  
\Phi_1^{ij} \equiv \Sigma(v^iv^j)
\,,
\nonumber \\
\Phi_1 &\equiv \Sigma(v^2) \,, \qquad
\Phi_2 \equiv \Sigma(U) \,, \qquad  X \equiv X(1) \,.
\label{eq:potentials1PN}
\end{align}

At 2PN order, these potentials also occur, along with specifically 2PN two-point functions,
\begin{align}
V_2^i &\equiv \Sigma(v^iU) \,, \qquad  \Phi_2^i \equiv \Sigma(V^i) \,,
\qquad X^i \equiv X(v^i)  \,,
\nonumber \\
X_1 &\equiv  X(v^2) \,, \qquad  X_2 \equiv X(U) \,, \qquad Y \equiv Y(1) \,,
\label{eq:potentials2PN}
\end{align}
where
\begin{equation}
Y(f) \equiv \int_{\cal M} {\rho^*(t,{\bm x}^\prime)f(t,{\bm
x}^\prime)}
{|{\bm x}-{\bm x}^\prime |^3 } d^3x^\prime 
\end{equation}
is a superduperpotential.

But a new kind of potential appears at 2PN order, called the triangle, or three-point potential, which depends on the field point and on {\em two} source points.  It is defined by the integral
\begin{align}
{\cal G}(A,B,C) &\equiv \frac{1}{4\pi} \int \frac{d^3 x'}{|{\bm x}_A - {\bm x}'||{\bm x}_B - {\bm x}'||{\bm x}_C - {\bm x}'|}
\nonumber 
\\
&= 
- \ln \Delta(ABC) + 1 \,,
\label{eq:calG}
\end{align}
where
\begin{equation}
\Delta(ABC) = |{\bm x}_A - {\bm x}_B|+|{\bm x}_A - {\bm x}_C|+|{\bm x}_B - {\bm x}_C| \,.
\end{equation}
This arises from Poisson potentials (Eq.\ (\ref{definepoisson})) of two two-point potentials.  Specific examples of triangle functions at 2PN order are
\begin{align}
P_2^{ij} &\equiv P(U^{,i}U^{,j}) \,, \qquad  P_2 \equiv
P_2^{ii}=\Phi_2
-{1 \over 2}U^2 
\,,\nonumber \\
G_1 &\equiv P({\dot U}^2)  \,, \qquad \qquad  G_2 \equiv P(U {\ddot U}) \,,
\nonumber \\
G_3 &\equiv -P({\dot U}^{,k} V^k) \,, \qquad  G_4 \equiv
P(V^{i,j}V^{j,i}) \,,\nonumber \\
G_5 &\equiv -P({\dot V}^k U^{,k}) \,, \qquad  G_6 \equiv P(U^{,ij}
\Phi_1^{ij}) \,,\nonumber \\
G_7^i &\equiv P(U^{,k}V^{k,i}) + {3 \over 4} P(U^{,i}\dot U ) \,.
\label{potentiallist}
\end{align}
  Also appearing at 2PN order is a quadrangle or four-point function $H$, which depends on $\bm x$ and on three source points.
It is defined by the integral
\begin{align}
{\cal H}& (A,B;C,D)\equiv \frac{1}{(4\pi)^2} \int \int \frac{d^3 x' d^3x''}{|{\bm x}' - {\bm x}''|}
\nonumber \\
\nonumber \\
&\quad \times 
\frac{1}{|{\bm x}_A - {\bm x}'||{\bm x}_B - {\bm x}'||{\bm x}_C - {\bm x}''||{\bm x}_D - {\bm x}''|}
\,.
\end{align}
Unfortunately, there is no known analytic form for this integral analogous to Eq.\ (\ref{eq:calG}).  The specific potential is
\begin{equation}
H \equiv P(U^{,ij} P_2^{ij}) \,.
\end{equation}

At 3PN order, the two-point potentials are joined by a megasuperpotential, 
\begin{align}
Z(f) &\equiv \int_{\cal M} {\rho^*(t,{\bm x}^\prime)f(t,{\bm
x}^\prime)}
{|{\bm x}-{\bm x}^\prime |^5 } d^3x^\prime\,.
\label{eq:definemega}
\end{align}
\begin{widetext}
The triangle potentials are joined by a triangle superpotential, defined by the integral
\begin{align}
{\cal F}(A,B;C) &\equiv  \frac{1}{4\pi} \int \frac{|{\bm x_C} - {\bm x}'| \,d^3 x'}{|{\bm x}_A - {\bm x}'||{\bm x}_B - {\bm x}'|}
\nonumber \\
& = \frac{1}{6} \biggl \{ r_{AC}r_{BC}-r_{AB}(r_{AC}+r_{BC})
-(r_{AC}^2+r_{BC}^2-r_{AB}^2)
\ln \Delta \biggr \} 
\nonumber \\
& \quad
 - \frac{1}{36} (r_{AC}^2+r_{BC}^2-r_{AB}^2)
-\frac{1}{6} r_C^2 
-\frac{1}{3} {\bf x}_C \cdot ( {\bf x}_{AC}+ {\bf
x}_{BC}) \,.
\end{align}
Here the field point could be ${\bm x}_A$ or ${\bm x}_C$.  Examples of $F$-type potentials are $P^{,i}(U^{,j} \ddot{X}^{,k})$ and $\ddot{S}^{,i}(U^{,j}U^{,k})$.

Additional quadrangle, or four-point potentials and superpotentials appear, such as 
\begin{align}
J(A,B,C,D)  &\equiv \frac{1}{4\pi} \int
\frac{d^3 x' }{|{\bf x}'-{\bf x}_A||{\bf x}'-{\bf x}_B|
|{\bf x}'-{\bf x}_C||{\bf x}'-{\bf x}_D|} \,,
\nonumber \\
K(A,B;C;D)  &\equiv \frac{1}{(4\pi)^2} \int \int
\frac{ |{\bf x}''-{\bf x}_D| \,d^3 x' d^3 x'' }
{|{\bf x}'-{\bf x}_A||{\bf x}'-{\bf x}_B||{\bf x}'-{\bf x}''|
|{\bf x}''-{\bf x}_C|} \,,
\nonumber \\
L(A,B;C,D)  &\equiv \frac{1}{(4\pi)^2}\int \int
\frac{ |{\bf x}'-{\bf x}''| d^3 x' d^3 x'' }{|{\bf x}'-{\bf x}_A||{\bf x}'-{\bf x}_B|
|{\bf x}''-{\bf x}_C||{\bf x}''-{\bf x}_D|} \,.
\nonumber
\end{align}
These also have no known analytic forms. Examples of these potentials are $P^{,i}(UU^{,k}V^{j,k})$ ($J$), 
$\ddot{S}^{,i}(U^{,jk} P_2^{jk})$ ($K$) and
$P^{,i}(U^{,jk} \ddot{X}^{,k})$ ($L$).
\end{widetext}

\subsection{Equations of motion for fluid balls}

We now address the equations of motion of fluid balls, and the appearance of structure-dependent effects at 3PN order.

In DIRE II \cite{2002PhRvD..65j4008P}, we defined the baryon rest mass, center of baryonic
mass, velocity and acceleration of each body by the formulae
\begin{eqnarray}
m_A &\equiv & \int_A \rho^* d^3x \,, 
\nonumber \\
{\bm x}_A & \equiv & (1/m_A) \int_A \rho^* {\bm x}^i d^3x \,, 
\nonumber \\
{\bm v}_A & \equiv & d{\bm x}_A /dt = (1/m_A) \int_A \rho^* {\bm v}^i d^3x \,, 
\nonumber \\
{\bm a}_A & \equiv & d{\bm v}_A /dt = (1/m_A) \int_A \rho^* {\bm a}^i d^3x \,, 
\label{rhostardefinitions}
\end{eqnarray}
where we  used the general fact, implied by the equation of continuity for
$\rho^*$, that 
\begin{align}
{\partial \over {\partial t}} \int \rho^*&(t, {\bm x}^\prime)  f(t, {\bm x},{\bm
x}^\prime) d^3x^\prime 
\nonumber \\
&= \int \rho^*(t, {\bm x}^\prime) \left ( {\partial
\over {\partial t}} + {\bm v}^\prime \cdot \nabla^\prime \right ) f(t, {\bm
x},{\bm x}^\prime) d^3x^\prime \,.
\label{continuity2}
\end{align}
With these definitions, we wrote the
coordinate acceleration  of the $A$-th body in the form
\begin{equation}
a^i_A = (1/m_A) \int_A \rho^* \frac{dv^i}{dt} d^3x \,,
\label{acceleration1}
\end{equation}
where $dv^i/dt$ is given by Eq.\ (\ref{eq:fluiddvdt}).
In this paper, we will be concerned only with the conservative 1PN, 2PN and 3PN terms listed in Appendix \ref{app:A}.  Note, however, that when time derivatives of a variable generate an acceleration, one must substitute the equations of motion valid to the appropriate order, 2PN order in 1PN terms, 1PN order in 2PN terms and Newtonian order in 3PN terms.

\subsection{Treatment of spherical ``pointlike'' masses}

We integrate all potentials that appear in the equation of motion,
as well as the equation of motion (\ref{eq:fluiddvdt}) itself over the bodies in
the binary system.  We treat each body as a non-rotating, spherically symmetric
fluid ball (as seen in its momentary rest frame), whose characteristic
size $s$ is much smaller than the orbital separation $r$.  We shall
discard all terms in the resulting equations that are proportional to
positive powers of $s$; these become smaller as the bodies become more compact.  The leading
Newtonian quadrupole 
effect is formally of order $(s/r)^2$ relative to the monopole gravitational
potential $m/r$, but for compact objects such as neutron stars or
black holes, $s \sim m$, so {\it effectively} this is comparable to a
2PN term.  Furthermore, if the quadrupole moment is the result of tidal
interactions with the companion, the size of the induced moment is of order
$(s/r)^3$, so the net effect is $O(s/r)^5$, or roughly 5PN order.  
Such leading multipolar terms can be calculated straightforwardly, including their PN corrections
but here we ignore them.  These terms are central to the concept of extracting information about the nuclear equation of state from the effects of tidal deformations in binary neutron star or neutron-star black-hole mergers.
Later, we will discuss the possible presence of terms proportional to positive powers of $s$ that are {\em not} related to conventional tides and that may occur before 5PN order.

We also discard all terms that are proportional to negative powers of
$s$: these correspond to ``self-energy'' corrections of PN and higher
order.   At 1PN order, when combined with suitable Newtonian virial theorems, such terms can be merged
uniformly into a suitably renormalized mass for each body, so that all internal structure dependence in the equations of motion that grows as the bodies shrink vanishes, in line
with the Strong Equivalence Principle (SEP) (see \cite{PW2014} for a pedagogical derivation).  Of course in alternative theories of gravity, this fails, leading to the well-known Nordtvedt effect \cite{tegp2}.   This also holds at 2PN order in general relativity \cite{2007PhRvD..75l4025M}: terms at 2PN order proportional to $s^{-1}$ and $s^{-5/2}$ vanished when absorbed into a 2PN-corrected mass for each body (see also \cite{1983SvAL....9..230G}).   It is not known if this holds at 3PN order, but we will not address that in this paper.

We keep only terms that are proportional to $s^0$.  For the most
part, these are the expected terms that depend on the two masses, terms
that
one would have obtained from a ``delta-function'' approach that
discarded all divergent self-energy terms.  However, at higher PN
orders, another class of $s^0$ terms is possible, at least in
principle.  These are terms that arise from non-linear combinations of
potentials.  Imagine the potential from the companion body being expanded in a multipolar expansion about  the center of mass of the body under study
in positive powers of $s$.  This could then be multiplied by another
potential which
 is a ``self-energy'' potential of that body, dependent upon negative
powers of $s$.  One could then end up with a term that has a piece that is
independent of the scale size $s$ of the body, but that still
depends on its internal density distribution.  In Sec.\ III.B of DIRE II, we showed that such
terms cannot appear at 1PN order by a simple symmetry argument (essentially gravity is not yet sufficiently non-linear to cause problems).  At
2PN order, terms of this kind {\it could} appear in certain non-linear
potentials, but in fact we showed (Sec.\ III.C and Appendix D of DIRE II) that they all  
vanish identically term by term by a subtler symmetry.    But already in DIRE II we pointed out that at 3PN order, that particular cancellation would no longer occur, leading potentially to $s^0$ contributions.

Our assumption that the bodies are non-rotating implies that
every element of fluid in the body has the same coordinate velocity,
so that $\bm v$ can be pulled outside any integral.    In principle one could let $\bm v$ within each body $A$ consist of a center-of-mass piece ${\bm v}_A$ (see Eq.\ (\ref{rhostardefinitions})) and an average-free, random piece ${\bar{\bm v}}$ to represent internal energy and pressure.  This assumption was used in our earlier proof of SEP to 2PN order  \cite{2007PhRvD..75l4025M}.  To keep things simple, we did not do that in our 3PN work.

Finally we assume that each body is suitably spherical.  By this we
mean that, in a local inertial frame comoving with
the body and centered at its baryonic center of mass, the baryon density
distribution is static and spherically symmetric in the coordinates of
that frame.  In Appendix B of  DIRE II, we showed that the transformation
between our global harmonic coordinates $x^i$ and the spatial coordinates
${\hat x}^i$ of this frame would generate relativistically
induced multipole moments for the body, albeit of 1PN and higher order
relative to the monopole moment.  Ordinarily, these would result in
terms of positive powers of $s$ in the equations of motion, which 
we ignore (in other words, as
the body's size shrinks to zero, the flattening become irrelevant);
however, as before, in terms with products of  potentials, we must worry
about the effect of self-potentials with negative powers of $s$ offsetting
the positive powers from the flattening. 
We showed  however that no $s^0$ terms arise in the
equations of motion at 2PN order from this effect, but that they {\em would} contribute in
principle at 3PN
order.

\section{Terms with $s^0$ scaling at 3PN order}
\label{sec3}

In this section we give some specific examples of 3PN terms that generate structure-dependent effects with $s^0$ scaling, and discuss the likelihood that these effects will miraculously cancel or be transformed away.  
Since we will be focusing on terms that are already of 3PN order, we can ignore relativistic flattening and  assume that the bodies are strictly spherical.
For a given body $A$, we will define $\bar{\bm x} \equiv {\bm x} - {\bm x}_A$, with $\bar{r} = |\bar{\bm x} |$. 

\subsection{Terms with triangle potentials}

We illustrate this with the boxed 3PN  term $U^{,i} P_2^{jk}$ in Eq.\ (\ref{eq:vjvk}), ultimately to be contracted with $v_1^j v_1^k$.  Expanding the potentials into  contributions from body 1 and from body 2, we obtain
\begin{align}
\frac{1}{m_1} &\int_1 \rho^* U^{,i} P_2^{jk} d^3x= \frac{1}{m_1} \int_1 \rho^*d^3x \left (U_1^{,i} + U_2^{,i} \right ) 
 \nonumber 
\\
& \quad \times \left (P_2^{jk}(1,1) + 2 P_2^{(jk)}(1,2)+P_2^{jk}(2,2) \right ) \,.
\label{eq:UPterm0}
\end{align}
Within body 1, under spherical symmetry, $P_2^{jk}(1,1)$ has the form (DIRE II, Eqs.\ (C6) and (C7))
\begin{equation}
P_2^{ij} \equiv  \tfrac{1}{3}\delta^{ij} P_2 + \bar{n}^{<ij>}PT_2 \,,
\label{P2split}
\end{equation}
where $\bar{n}^{<ij>} = \bar{n}^{ij} - \delta^{ij}/3$.  The trace $P_2$ and the traceless part $PT_2$
satisfy the equations $\nabla^2 P_2 = - (U^\prime)^2$, and 
$\nabla^2 PT_2 -6PT_2/{\bar r}^2 = -(U^\prime)^2$, where $U' = dU/d{\bar r}$, such that $U^{,i} = \bar{n}^i U'$.  Note that $P_2$ and $PT_2$ scale as $(m_1/s)^2$

The potential $P_2^{jk}(1,2)$ has the form, within body 1,
\begin{align}
P_2^{jk}(1,2) &= m_2 \int_1 {\rho^*}' d^3x' \left [ \frac{ (\hat{y}_1-\hat{y}'_{2})^j (\hat{y}_2+\hat{y}'_{2})^k}{\Delta^2}
\right .
\nonumber \\
& \left . \qquad \qquad
+ \frac{(\delta^{jk} - {\hat{y'}}_{2}^j  \hat{y'}_{2}^k)}{y_2 \Delta} \right ] \,,
\end{align}
where $y_1^j = (x-x')^j$, $y_2^j = (x-x_2)^j$, ${y'}_2^{j} = (x' -x_2)^j$, $\Delta =
y_2 + y'_2 + y_1$, and hats denote the corresponding unit vector. Here, body 2 can be treated as a point mass.

Finally, $P_2^{jk}(2,2)$ at body 1 has the form
\begin{equation}
P_2^{jk}(2,2) = \frac{m_2^2}{4 y_2^2} \left ( \hat{y}_2^{jk} - \delta^{jk} \right ) \,.
\end{equation}

We also make use of the expansions of $U_2$ and its gradient about the center of mass of body 1, 
\begin{align}
U_2 &= \sum_\ell \frac{m_2}{\ell !} \bar{x}^L \partial_L \frac{1}{r} \,,
\nonumber \\
\partial^i U_2 &= \sum_\ell \frac{m_2}{\ell !} \bar{x}^L \partial_{iL} \frac{1}{r} \,,
\label{eq:U2expand}
\end{align}
where $\bar{x}^j = (x - x_1)^j$ and $r = |{\bm x}_{12}|$.

The term $U_1^{,i} P_2^{jk}(1,1)$ in Eq.\ (\ref{eq:UPterm0}) vanishes because it involves only body 1, and contains an odd number of unit vectors, whose angular integral over the spherical body must vanish.  For the term $U_2^{,i} P_2^{jk}(1,1)$, since $P_2^{jk}(1,1)$ scales as $s^{-2}$, we need to keep only the term in the expansion of $U_2^{,i}$ in Eq.\ (\ref{eq:U2expand}) that scales as $s^2$. The result is a term
\begin{align}
&\frac{m_2}{2m_1} \int_1 \rho^* \bar{r}^2 n^{lm} \partial^{ilm} (r^{-1}) \left (\tfrac{1}{3} \delta^{jk} P_2 + n^{\langle jk \rangle} PT_2 \right ) d^3x
\nonumber \\
& \quad=- \frac{m_2}{2m_1} \int_1 \rho^* \bar{r}^2 PT_2 d^3x (\delta^{lm} \delta^{jk} + {\rm sym} )  \frac{N^{\langle ilm \rangle}}{r^4}
\nonumber \\ 
& \quad =-\frac{m_2 m_1^2}{r^4} \eta_3 N^{\langle ijk \rangle} \,,
\end{align}
where $+ {\rm sym}$ denotes the terms needed to make the expression completely symmetric on its indices, and where 
\begin{equation}
\eta_3 \equiv \frac{1}{m_1^3} \int_1 \rho^* \bar{r}^2 PT_2 d^3x 
\end{equation}
is a dimensionless coefficient that scales as $s^0$, i.e.\ is independent of the compactness of the body.  This and many other structure coefficients will be discussed in Sec.\ \ref{sec:structurecoeffs}.

For the term $U_1^{,i} P_2^{jk}(1,2)$, we need to expand $P_2^{jk}(1,2)$ in powers of $s$, keeping only $s^2$ terms to cancel the $s^{-2}$ scaling from $U_1^{,i}$, and also keeping only odd numbers of unit vectors within body 1 to go along with the single unit vector from $U_1^{,i}$.  It is useful to write 
\begin{align}
y_2^j &= x_{12}^j + \bar{x}^j \,,
\nonumber \\
{y'}_2^{j} &=  x_{12}^j + \bar{x}^j - y_1^j \,,
\end{align}
and to treat $\bar{x}^j$ and $y_1^j$ as of order $s$ in the expansion (the unit vector $\hat{y}_1^j $ is treated as $O(s^0)$).  The result is
\begin{align}
&\frac{5m_2}{4m_1}\frac{N^{\langle mkl \rangle}}{r^4}  \int_1 \rho^* n^i U' d^3x \int_1  {\rho^*}' d^3x' y_1^{-1} \times
\nonumber \\
&\quad  \left [  \left (2 y_1^l  - 3  \bar{r} n^l \right ) y_1^2 \delta^{mj}
+ y_1^{jlm} - 3 y_1^{jl} \bar{r} n^l + 3 y_1^j \bar{r}^2 n^{lm}  \right ] \,.
\end{align} 
We now make use of a number of identities, such as
\begin{align}
y_1^j y_1 &= \tfrac{1}{3} \partial^j y_1^3 \,,
\nonumber \\
y_1^{jk}/y_1 &= \tfrac{1}{3} \partial^{jk} y_1^3 - \delta^{jk} y_1 \,,
\nonumber \\
y_1^{jkl}/y_1 &= \tfrac{1}{15} \partial^{jkl} y_1^5 - \delta^{(jk} \partial^{l)} y_1^3 \,.
\end{align}
When integrated over ${\rho^*}' d^3x' $, the quantities $y_1$, $y_1^3$ and $y_1^5$ yield the superpotentials $X$, $Y$ and $Z$ (see Sec.\ \ref{sec:zoo} for definitions), which satisfy $\nabla^2 X = 2U$, $\nabla^2 Y = 12 X$ and $\nabla^2 Z = 30 Y$.  For spherically symmetric functions $f(r)$, we also have the useful identities 
\begin{align}
\partial^{jk} f(r) &= n^{jk} \nabla^2 f - \frac{3}{r} n^{\langle jk \rangle} f' \,,
\nonumber \\
\partial^{jkl} f(r) &= n^{jkl}\left ( \nabla^2 f  \right )'- \frac{5}{r} n^{\langle jkl \rangle} \left ( \nabla^2 f- \frac{3f'}{r^2} \right )\,.
\end{align}
Making use of these identities and integrating over the angular part of $d^3x$ in body 1, we obtain
the result
\begin{align}
&\frac{m_2}{m_1}\frac{N^{\langle ijk \rangle}}{r^4}  \int_1 \rho^* U' \left ( \tfrac{5}{12} Y' - 2\bar{r} X + \tfrac{1}{2} \bar{r}^2 X' \right ) d^3x
\nonumber \\
&\quad  = - \frac{m_1^2 m_2}{12 r^4} N^{\langle ijk \rangle} \left (24\alpha_3 - 6\alpha_4 -5\beta_4 \right ) \,.
\end{align}
where $\alpha_3$, $\alpha_4$, $\beta_4$ are dimensionless structure coefficients defined by
\begin{align}
\alpha_3 &\equiv \frac{1}{m_1^3} \int_1 \rho^* U' X \bar{r} d^3x \,,
\nonumber \\
\alpha_4 &\equiv \frac{1}{m_1^3} \int_1 \rho^* U' X' \bar{r}^2 d^3x \,,
\nonumber \\
\beta_4 &\equiv \frac{1}{m_1^3} \int_1 \rho^* U' Y' d^3x \,.
\end{align} 
The final result for this term is automatically symmetric on $(jk)$.  The term 
$U_2^{,i} P_2^{jk}(1,2)$ has only point-mass contributions plus terms that scale as positive powers of $s$, so the result is 
\begin{equation}
-\frac{m_1^2 m_2}{2r^4} N^i (\delta^{jk} -2 N^j N^k ) \,.
\end{equation}
The term $U_1^{,i} P_2^{jk}(2,2)$ vanishes: $U_1^{,i}$ scales as $s^{-2}$, while the $s^2$ term in the expansion of $P_2^{jk}(2,2)$ has an even number of unit vectors $n^j$, so the angular integration vanishes.  The final term $U_2^{,i} P_2^{jk}(2,2)$ yields $(m_2^3/4r^4)N^i (\delta^{jk} - N^j N^k)$.  Combining all the terms and contracting with $v_1^j v_1^k$ yields the contribution to the acceleration
\begin{align}
a_1^i & = \frac{m_2^2 N^i}{r^4} \left [ m_2 \left (v_1^2 - Nv_1^2 \right ) - 4m_1  \left (v_1^2 - 2 Nv_1^2 \right ) \right ]
\nonumber \\
& \quad + \frac{2}{15} \frac{m_1^2 m_2}{r^4} \left (24\alpha_3 - 6\alpha_4 -5\beta_4 + 6\eta_3 \right ) 
\nonumber \\
& \qquad \quad \times \left [ N^i (v_1^2 -5 Nv_1^2) - v_1^i Nv_1 \right ] \,,
\label{eq:P2terminEOM}
\end{align}
where $Nv_1 = {\bm N} \cdot {\bm v}_1$.
Interestingly, in the list of explicitly 3PN terms in Eqs.\ (\ref{eq:viv2}) - (\ref{eq:vjvk}), this is the first to exhibit structure-dependent effects.  While triangle potentials appear in Eqs.\ (\ref{eq:viv2}), (\ref{eq:vivj}) and (\ref{eq:v2}) (as well as in (\ref{eq:vi})),  they are basically 2PN potentials, which were shown in DIRE II  to have no structure dependent effects, multiplied by factors of velocity.  For example, because the structure-dependent effects in $U_2^{,i} P_2^{jk}$ are proportional to $N^{\langle ijk \rangle}$, the contraction of the indices in the term $v^2 U^{,j} P_2^{ij}$ in Eq.\ (\ref{eq:v2}) kills them.   Triangle potentials tend to generate structure-dependent terms when they are paired with a two-body potential.  The potential can be multiplicative, as in the example just analyzed, or internal, as in term 18 of Eq.\ (\ref{eq:vjvk}), involving  $P^{,i}(\Phi_2^{,(j}U^{,k)})$.  Time derivatives can generate accelerations, which then produce internal potentials, as in term 51 of  Eq.\ (\ref{eq:vj}), involving $P^{,[i}(U^{,j]} \dot{\Phi}_1)$.  The ${\cal F}$ triangle superpotential  makes its first appearance in the last four terms of Eq.\ (\ref{eq:vjvk}), and repeatedly thereafter, and generates structure-dependent terms via the numerous time derivatives.

\subsection{Highly non-linear two-point terms}

To illustrate that even the seemingly simplest term can lead to $s^0$ effects, we evaluate the first 3PN term with no explicit velocity dependence in Eq.\ (\ref{eq:nov}), the boxed term $-(32/3) U^3 U^{,i}$.  It does not involve triangle potentials.
We expand the potential into contributions from body 1 and from body 2, giving
\begin{align}
\frac{1}{m_a} \int_1 \rho^* U^{,i} U^3 d^3x&= \frac{1}{m_a} \int_1 \rho^*d^3x \left (U_1^{,i} + U_2^{,i} \right ) 
\nonumber \\
& \quad \times \left ( U_1^3  + 3U_1^2 U_2 + 3U_1 U_2^2 + U_2^3 \right ) \,.
\end{align}
The integral involving $U_1^{,i} U_1^3 $ is purely internal to body 1 and vanishes by spherical symmetry.  In the term $U_1^{,i} U_1^2 U_2$, there are $\ell$ unit vectors from $n^L$ in the expansion of $U_2$ and one from the gradient of $U_1$, so by symmetry, the integral will vanish unless $L$ is odd.  The combination $U_1^{,i} U_1^2$ scales as $s^{-4}$, while $\bar{x}^L$ scales as $s^\ell$, so the  integrand will then have the scaling $s^{\ell - 4} \sim s^{-3},\, s^{-1},\, s^{+1}, \dots$, hence no $s^0$ term.  A similar argument gives no $s^0$ contribution from the combination $U_2^{,i} U_1^3$.
In the term $U_1^{,i} U_2^3$, there are $\ell + \ell' + \ell''$ unit vectors from the product of the $U_2$s and one from the gradient of $U_1$, so the integral will vanish unless $\ell + \ell' + \ell''$ is odd.  The scaling will then be $s^{\ell + \ell' + \ell''-2} \sim s^{-1},\, s^{+1}, \dots$, hence no $s^0$ term.  The same argument applies to the term  $U_2^{,i} U_1 U_2^2$.  The combination $U_2^{,i} U_2^3$ gives the expected point mass result, $-m_2^4 n_i /r^5$, plus terms scaling as positive powers of $s$. 

The offending terms are $U_1^{,i} U_1 U_2^2$ and $U_2^{,i} U_1^2 U_2$.  The first term has $\ell + \ell'$ unit vectors from $U_2^2$ and one from $U_1^{,i}$, so $\ell + \ell'$ must be odd, implying a scaling of $s^{\ell + \ell'-3} \sim s^{-2},\, s^0,\, s^{+2} \dots$. The scaling $s^0$ will therefore occur when $(\ell, \ell') = (3,0),\, (0,3) ,\, (1,2),\, (2,1)$.
The integral becomes
\begin{align}
\frac{1}{m_1} &\sum_{\ell \ell'} \frac{m_2^2}{\ell! \ell'!} \partial_L \frac{1}{r} \partial_{L'} \frac{1}{r} \int_1 \rho^* {\bar r}^{\ell + \ell'} {\hat n}^{L+L'+1} U'_1 U_1 d^3x
\nonumber \\
=&\frac{m_2^2}{ 15m_1} \int_1 \rho^* {\bar r}^3  U'_1 U_1 d^3x
 \left ( \delta^{ij} \delta^{kl} + \delta^{ik} \delta^{jl} +\delta^{il} \delta^{jk} \right )
 \nonumber \\
 & \times \left [ \frac{1}{3r} \partial^{jkl}  \left(\frac{1}{r}\right ) +
  \partial^j  \left(\frac{1}{r}\right ) \partial^{kl}  \left(\frac{1}{r}\right ) \right ]
  \nonumber \\
  =& -\frac{4}{15} \frac{m_1^2 m_2^2}{r^5} N^i \zeta_2 \,.
\end{align} 
We use the fact that $\nabla^2 (1/r) = 0$, since $r = r_{12}$, and we define the dimensionless coefficient
\begin{equation}
\zeta_2 \equiv \frac{1}{m_1^3} \int_1 \rho^* {\bar r}^3 U' U d^3x \,.
\end{equation}
In the second term, there are $\ell + \ell'$ unit vectors from $U_2^{,i} U_2$, so $\ell + \ell'$ must be even, which implies the scaling $s^{\ell + \ell'-2} \sim s^{-2},\, s^0,\, s^{+2}, \dots$, and therefore that $(\ell, \ell') = (0,2),\,(2,0),\, (1,1)$.  The result is
\begin{align}
\frac{m_2^2}{ 3m_1}& \int_1 \rho^* {\bar r}^2 U_1^2 d^3x \delta^{jk}
 \left [ \frac{1}{2r} \partial^{ijk}  \left(\frac{1}{r}\right ) 
\right .
\nonumber \\
& \left . \quad +
 \frac{1}{2}\partial^{i}  \left(\frac{1}{r}\right ) \partial^{jk}  \left(\frac{1}{r}\right )
 + \partial^k \left(\frac{1}{r}\right ) \partial^{ij}  \left(\frac{1}{r}\right ) \right ]
 \nonumber \\
  =& -\frac{2}{3} \frac{m_1^2 m_2^2}{r^5} N^i \zeta_1 \,,
\end{align}
where 
\begin{equation}
\zeta_1 \equiv \frac{1}{m_1^3} \int_1 \rho^* {\bar r}^2 U^2 d^3x \,.
\end{equation}
The contribution of this 3PN term to the equation of motion is
\begin{equation}
a_1^i = \frac{32}{3} \frac{m_2 n^i}{r^2} \left [ \left (\frac{m_2}{r} \right )^3 + \frac{2m_1^2 m_2}{5r^3} \left ( 5 \zeta_1 + 2 \zeta_2 \right ) \right ] \,.
\label{eq:UcubedtermEOM}
\end{equation}

\begin{table*}[t]
\caption{Structure coefficients (an incomplete list)}
\begin{tabular}{|l@{\hskip 0.2 cm} | l@{\hskip 1 cm} || l@{\hskip 0.2 cm} | l @{\hskip 1 cm}|| l@{\hskip 0.2 cm} | l @{\hskip 1 cm}|| l@{\hskip 0.2 cm} | l @{\hskip 1 cm}|}
\hline
Coeff&$f(r)$&Coeff&$f(r)$&Coeff&$f(r)$&Coeff&$f(r)$\\
\hline
&&&&&&&\\
$\alpha_1$&$UX$&
$\beta_1$&$UY/r^2$&
$\gamma_4$&$U^\prime Z^\prime /r^2$&
$\zeta_1$&$U^2r^2$
\\
$\alpha_2$&$UX^\prime r$&
$\beta_2$&$UY^\prime /r$&
$\gamma_6$&$U^{\prime\prime} Z^\prime /r$&
$\zeta_2$&$UU^\prime r^3$
\\
$\alpha_3$&$U^\prime Xr$&
$\beta_3$&$U^\prime Y/r$&
&&
$\zeta_4$&$U^\prime U^\prime r^4$
\\
$\alpha_4$&$U^\prime X^\prime r^2$&
$\beta_4$&$U^\prime Y^\prime $&
$\lambda_2$&$rF_2^\prime$&
$\zeta_5$&$U^{\prime\prime}U r^4$
\\
$\alpha_5$&$U^{\prime\prime} Xr^2$&
$\beta_5$&$U^{\prime\prime} Y $&
$\lambda_3$&$FT_2$&
$\zeta_6$&$U^{\prime\prime}U^\prime r^5$
\\
$\alpha_6$&$U^{\prime\prime} X^\prime r^3$&
$\beta_6$&$U^{\prime\prime} Y^\prime r$&
$\lambda_4$&$rFT_2^\prime$&
&
\\
&&&&&&$\pi_1$&$rF_5$
\\
$\eta_1$&$r^2 P_2$&
$\omega_3$&$r^2 PT_3$&
$\tau_1$&$F_4$&
$\pi_2$&$r^2F_5^\prime$
\\
$\eta_2$&$r^3 P_2^\prime $&
$\omega_4$&$r^3 PT_3^\prime $&
$\tau_2$&$rF_4^\prime$	&
$\pi_3$&$rFS_5$
\\
$\eta_3$&$r^2 PT_2$&
&&
$\tau_3$&$FT_4$&
$\pi_4$&$r^2FS_5^\prime$
\\
$\eta_4$&$r^3 PT_2^\prime $&
$\mu_2$&$r {\cal G}^\prime$&
$\tau_4$&$rFT_4^\prime$	&
$\pi_5$&$rFT_5$
\\
&&&&&&$\pi_6$&$r^2FT_5^\prime$\\
\hline
\end{tabular}
\label{tab:coeffs}
\end{table*}

\subsection{Terms with quadrangle potentials}

Terms involving the quadrangle potentials and superpotentials $\cal H$, $\cal J$, $\cal K$, and $\cal L$ are the most difficult to deal with because of the lack to date of any general analytic expressions beyond the simplest two-body case.  Using a tailor-made and cumbersome method \cite{2007PhRvD..75l4025M}, we showed that the 2PN quadrangle term involving $H^{,i}$ in Eq.\ (\ref{eq:2PN}) produced only $s^{-1}$ structure dependent terms (which combined with others to validate the SEP at 2PN order), but no $s^0$ terms.  The same holds true for simple 3PN terms where $H^{,i}$ appears multiplied by velocities, as in the final terms of Eqs.\ (\ref{eq:vivj}) and (\ref{eq:v2}), or is hit with a time derivative, as in the final term of Eq.\ (\ref{eq:vi}).  Apart from these simple cases, more complex ${\cal H}$  potentials appear in terms 80 - 84 of  Eq.\ (\ref{eq:vj}), involving Poisson potentials of triangle potentials, and 
${\cal J}$ potentials appear in terms 85 - 87, involving Poisson potentials of three two-body potentials.  In the terms without explicit velocity dependence, Eqs.\ (\ref{eq:nov}), the full array of $\cal H$, $\cal J$, $\cal K$, and $\cal L$ potentials appears, beginning with the term $UH^{,i}$ (boxed) through the remaining terms in that equation.  

It turns out that one needs analytic expressions for the terms involving quadrangle potentials {\em only} for the restricted case of three bodies, since we care about one body (treated as a point mass) and two points within the other body.    To this end, an early paper by Schaefer  
\cite{1987PhLA..123..336S} may be useful: there the author obtained the contribution of the quadrangle potential to the three-body Lagrangian to 2PN order.  Whether the techniques used in that paper can be applied to the many versions of the $\cal H$, $\cal J$, $\cal K$, and $\cal L$ potentials is an open question.

\subsection{The structure coefficients}
\label{sec:structurecoeffs}

At the time the project was halted, many terms had been found with structure dependent coefficients scaling as $s^0$.   The number of distinct coefficients was large, and seemed to grow with each new class of terms analyzed.  Table \ref{tab:coeffs} lists the coefficients defined and recorded at that time.  If the notation seems somewhat scattershot, it is because the symbols were assigned on the fly.    For a given function $f(r)$, the coefficient is given by 
\begin{equation}
{\rm Coefficient} = \frac{1}{m_1^3} \int_1 \rho^* f(r) d^3x \,.
\end{equation}
There are cases of relations connecting the coefficients, including $2\eta_3 + 2\omega_3 + 3\mu_2 + \zeta_1 = 0$ but these are rare, and unlikely to control the growing number and type of coefficients.

In addition to $U$, $X$, $Y$ and $Z$ defined earlier, the functions shown in Table \ref{tab:coeffs} are defined (all assuming spherical symmetry) by
\begin{align}
\nabla^2 P_2 &= -(U^\prime)^2 \,,
\nonumber \\
\nabla^2 PT_2 - 6PT_2/r^2 &= -(U^\prime)^2 \,,
\nonumber \\
\nabla^2 P_3 &= 4\pi \rho^* U \,,
\nonumber \\
\nabla^2 PT_3 - 6PT_3/r^2 &= -U (U^{\prime\prime}-U^{\prime}/r) \,,
\nonumber \\
\nabla^2 F_2 &= 2P_2 \,,
\nonumber \\
\nabla^2 FT_2 - 6FT_2/r^2 &= 2PT_2 \,,
\nonumber \\
\nabla^2 F_3 &= 2P_3 \,,
\nonumber \\
\nabla^2 FT_3 - 6FT_3/r^2 &= 2PT_3  \,,
\nonumber \\
\nabla^2 F_4 &= -U^\prime X^\prime \,,
\nonumber \\
\nabla^2 FT_4 - 6FT_4/r^2 &= -U^\prime X^\prime \,,
\nonumber \\
\nabla^2 F_5 -2 F_5 /r^2 &= X^\prime U^\prime/r \,,
\nonumber \\
\nabla^2 FS_5 - 2FS_5/r^2 &= X^\prime (U^{\prime\prime}-U^\prime/r)\,,
\nonumber \\
\nabla^2 FT_5 - 12FT_5/r^2 &= X^\prime (U^{\prime\prime}-U^\prime/r) \,,
\nonumber \\
\nabla^2 {\cal G} &= -U^2 \,.
\end{align}

Table \ref{tab:structcoeffvalues} shows values of the coefficients that appear in Eqs.\ (\ref{eq:P2terminEOM}) and
(\ref{eq:UcubedtermEOM}) for a uniform density body ($\rho^*=$ constant), an $n=1$ polytrope ($\rho^* = \rho^*_c \sin \xi / \xi$, $\xi = \pi r/R$) and a centrally condensed body ($\rho^* = \rho^*_c (1-\xi^2)^6$, $\xi = r/R$).    The coefficients do not vary strongly over these profiles, in large part because a centrally condensed density profile can be approximated as a uniform density body of a smaller radius, yet these coefficients are all independent of radius.  As a result, the change in the values of the coefficients between density profiles $(1-\xi^2)^6$ and $(1-\xi^2)^{75}$ is only a few percent.  The key point is that these coefficients are of order unity.

\begin{center}
\begin{table}[t]
\caption{Values of selected structure coefficients}
\begin{tabular}{c@{\hskip 0.5 cm}c@{\hskip 0.5 cm}r@{\hskip 0.5 cm}r@{\hskip 0.5 cm}r}
\hline
&&Uniform&Polytrope&Centrally\\
&$f(r)$&density&$n=1$&condensed\\
\hline
\\
$\zeta_1$&$U^2r^2$&0.79048&0.74229&0.71758
\\
$\zeta_2$&$UU'r^3$&-0.47619& -0.46592& -0.45996
\\
$\alpha_3$&$U'Xr$&-0.64762& -0.65321& -0.65994
\\
$\alpha_4$&$U'X'r^2$&-0.36190& -0.36569&-0.36695
\\
$\beta_4$&$U'Y'$&-2.28571& -2.29924& -2.32187
\\
$\eta_3$&$PT_2 r^2$&0.04048&0.04112&0.04200
\\
\hline
\end{tabular}
\label{tab:structcoeffvalues}
\end{table}
\end{center}

\bigskip 
\section{Discussion and Conclusions}
\label{sec:conclusions}

In this paper, we have tried to show that structure dependent effects of a particularly strange kind -- that do not scale with the size of the body -- could be present in the 3PN equations of motion of general relativity.  The crucial question, which remains unanswered, is: do these effects persist in the equations of motion, and subsequently in the gravitational waveforms, or do they magically disappear?  At the beginning of the project, there was a hope for a grand cancellation based upon our belief in the Strong Equivalence Principle, but the persistent growth in the number and kind of distinct coefficients with no evidence of significant links between them made this hope hard to sustain.  This also makes it difficult to imagine that, in combination perhaps with a change of variables for the position of each body, the coefficients might combine to produce a structure-{\em independent} factor related to the ambiguity coefficient of the MPPM approach.  

A criticism of our approach is that we have not included terms and potentials related to the pressure forces required to keep our bodies from collapsing, either by treating each velocity as a sum of a center-of-mass velocity and a random ``thermal'' velocity (as was done at 2PN order in  \cite{2007PhRvD..75l4025M}) or by including explicit terms and potentials generated by pressure $p$. Adding such effects makes the calculations even more complicated, and it is again not clear that such additions have the breadth to deal with the growing number of distinct coefficients.

If these structure-dependent effects are real, the question is will their neglect in gravitational-wave templates matter?  As can be seen in Eqs.\ (\ref{eq:P2terminEOM}) and (\ref{eq:UcubedtermEOM}) and in Table \ref{tab:structcoeffvalues}, the structure coefficients could alter the 3PN coefficients in the binary equations of motion by as much as 100 percent.  At current levels of GW sensitivity, the data are not sufficiently accurate to rule out such differences from the standard ``point-mass'' waveform coefficients; the bounds on deviations from the 3PN waveform parameters from analyses of data from the third observing run of LIGO-Virgo are at roughly the 50 percent level  \cite{2021arXiv211206861T}.  But with next-generation ground-based detectors promising up to a 10-fold improvement in amplitude sensitivity  \cite{2021PhRvD.103d4024P,2023arXiv230613745E}, the presence or absence of these effects could begin to matter.  In addition, any uncertainty regarding the existence of these effects could introduce an uncertainty in efforts to use 5PN tidal effects to elucidate the nuclear equation of state in double neutron star inspirals  \cite{2023ApJ...955...45F}.  

On the other hand, if structure-dependent effects occur at 3PN order, two PN orders before tidal effects,  could they represent an opportunity for testing neutron star structure?  The relative insensitivity of the values of the structure coefficients to the equation of state, as illustrated in Table \ref{tab:structcoeffvalues}, could reduce the effectiveness of such tests.  

In this paper, we have focused on $s^0$ effects, dropping terms with negative or positive powers of $s$.  For example, we discarded numerous terms proportional to $s$.  Assuming that they scale as $s/r$ relative to the 3PN term from which they came, then because $s \sim Gm/c^2$ they would represent effectively 4PN terms which depend {\em directly} on the radius of the neutron star, and are 1PN order below tidal effects.  If these are real, they might represent an alternative or complementary way to explore the internal structure of neutron stars.  In fact, even at 2PN order, such $s/r$ terms arose, but were discarded immediately; we plan to return to this and check whether such terms survive or whether they disappear via contraction of an STF tensor on a pair of its indices in the same way that 2PN $s^0$ terms disappeared (see the discussion following Eq.\ (\ref{eq:P2terminEOM})).   

The goal of this paper is to revive interest in this problem, which could be crucial for the future success of gravitational-wave science, particularly for sources containing material objects like neutron stars.  With modern tools based on machine learning or artificial intelligence it may be possible to complete the calculation more quickly and to deal with the more complicated quadrangle gravitational potentials, which at the time were an additional obstacle to completion of the work.
Conversely, if such calculations showed that such $s^0$ terms actually cancelled or were suitably removable, the result would be a remarkable validation of the Strong Equivalence Principle of general relativity.

\medskip
\acknowledgments

The work carried out at Washington University was supported in part by the National Science Foundation,
Grants No.\ PHY 03-53180 and PHY 06-52448, and by the National Aeronautics and Space Administration, Grant No. NNG06GI60G.
We want to acknowledge the hard work carried out during 2003-4 by graduate students Thomas Mitchell, Han Wang and Jing Zeng, and \'Ecole Normale Sup\'erieure student Emmanuelle Gouillart.  When the project was halted, each graduate student turned to a related topic in PN binary dynamics to complete the PhD \cite{2007PhRvD..75f4017W,2007PhRvD..75l4025M,2007GReGr..39.1661Z}, and Ms. Gouillart reported on her contributions to the project to complete her ENS internship.  Over the years, we had numerous discussions with Luc Blanchet, Guillaume Faye, Thibault Damour and others about these problematic structure dependent terms, with no clear resolution.  We are particularly grateful to Eric Poisson for showing us preliminary work with Tristan Pitre, and for encouraging us to (finally) write this paper.  
This paper was supported in part by NSF Grant No. PHY 22-07681.   

\begin{widetext}
\appendix

\section{Equations of motion to 3PN order in terms of $\rho^*$}
\label{app:A}

The 1PN and 2PN equations of motion have the form
\begin{eqnarray}
a_{PN}^i &=&
	v^2 U^{,i} -4 v^i v^j U^{,j} - 3v^i \dot U - 4 U U^{,i} 
	+ 8 v^j V^{[i,j]} 
	+ 4 \dot V^i +
\tfrac{1}{2} \ddot X^{,i} + \tfrac{3}{2} \Phi_1^{,i}
	-\Phi_2^{,i} \,,
	\label{eq:1PN}
	 \\
a_{2PN}^i &=&
 4 v^i v^j v^k V^{j,k} +  v^2 v^i \dot U  
+ v^i v^j ( 4 \Phi_2^{,j} - 2 \Phi_1^{,j} - 2 \ddot X^{,j} )
- \tfrac{1}{2} v^2 (2 \Phi_2^{,i} + \Phi_1^{,i} -  \ddot X^{,i} )
\nonumber \\
&&
+ v^j v^k \left ( 2 \Phi_1^{jk,i} - 4 \Phi_1^{ij,k} + 2 P_2^{jk,i} -4 P_2^{ij,k} \right)
+v^i \left ( 3 \dot \Phi_2 - \tfrac{1}{2} \dot \Phi_1 
	+4 V^k U^{,k} - \tfrac{3}{2} \dddot{X} \right)
\nonumber \\
&&
+ v^j  \left ( 
	8 V^i U^{,j} 
	- 16 U V^{[i,j]} 
        + 8 V_2^{[i,j]} 
	- 16 \Phi_2^{[i,j]} 
	- 4 \Sigma^{,[i}(v^{j]}v^2) 
	- 4 \dot \Phi_1^{ij} 
	+ 4 \ddot X^{[i,j]} 
	+ 32 G_7^{[i,j]}
	- 4 \dot P_2^{ij} \right )
\nonumber \\
&&
	+ 8 U^2 U^{,i} 
	- 6 U \Phi_1^{,i} 
	+ 4 U \Phi_2^{,i} 
	-2 U^{,i} \Phi_1 
	+ 4 U^{,i}\Phi_2 
	- 4 \Phi_1^{ij} U^{,j} 
	+ 8 V^j V^{j,i} 
	+ 4 V^i \dot U 
	- 8 U \dot V^i 
	- 8 \dot \Phi_2^i 
	+ 4 \dot V_2^i
\nonumber \\
&& 
	+ \tfrac{7 }{ 8} \Sigma^{,i}(v^4) 
	+ \tfrac{9 }{ 2} \Sigma^{,i}(v^2 U)
	+ \tfrac{3 }{ 2} \Sigma^{,i}(U^2) 
	- \tfrac{3 }{ 2} \Sigma^{,i}(\Phi_1) 
	- \Sigma^{,i}(\Phi_2) 
	- 4 \Sigma^{,i}(v^j V^j) 
	+ 2 \dot \Sigma (v^i v^2)
	-2 U \ddot X^{,i} 
	- 2 U^{,i}\ddot X 
\nonumber \\
&&
	+ 2 \stackrel{(3)}{X^{i}}
	- \tfrac{1}{2} \Sigma^{,i}(\ddot X )
	+ \tfrac{3 }{ 4} \ddot X_1^{,i} 
	- \tfrac{1}{2} \ddot X_2^{,i}
	+ \tfrac{1 }{ 24} \stackrel{(4)}{Y^{,i}} 
	- 6G_1^{,i} 
	- 4 G_2^{,i} 
	+ 8 G_3^{,i} 
	+ 8 G_4^{,i} 
	- 4 G_6^{,i} 
	+ 16 \dot G_7^i 
	- 4U^{,j} P_2^{ij} 
	- 4 H^{,i} \,.
	\nonumber
	\\
	\label{eq:2PN}
\end{eqnarray}

We organize the 3PN terms according to the number and type of explicit factors of velocity that occur: 

\begin{eqnarray}
a^i_{3PN}[v^3]&=& v^iv^jv^k \left (
 8UV^{k,j}
-8V^kU^{,j}
+2\Phi_{1,t}^{jk}
+2\ddot{X}^{k,j}
-8\Phi_{2}^{k,j}
+4V_{2}^{k,j}
+2V_{3}^{k,j}
+2\dot{P}_{2}^{jk}
+16G_{7}^{k,j}
\right ) 
\nonumber \\
%
&& + v^iv^2 \left (
4U\dot{U}
+4V^jU_{,j}
-\tfrac{1}{2}\dot{\Phi}_{1}
-\dot{\Phi}_{2}
+ \tfrac{1}{2}\dddot{X}
\right ) \,,
\label{eq:viv2}
\\
a^i_{3PN}[v^{ij}] &=& 
v^iv^j \left (
-8V^j\dot{U}
-16V^kV^{j,k}
-\ddot{X}_{1}^{,j}
+2\ddot{X}_{2}^{,j}
-\tfrac{1}{6}\stackrel{(4)\,\,}{Y^{,j}} 
-\tfrac{3}{2}\Sigma^{,j}(v^4)
-6\Sigma^{,j}(Uv^2)
-6\Sigma^{,j}(U^2)
\right .\nonumber \\
&&
\left .
+6\Sigma^{,j}(\Phi_1)
+4\Sigma^{,j}(\Phi_2)
+2\Sigma^{,j}(\ddot{X})
+12G_{1}^{,j}
+16G_{2}^{,j}
-32G_{3}^{,j}      
-16G_{4}^{,j}
+16G_{6}^{,j}
+16H^{,j}      
\right ) ,
\label{eq:vivj}
\\
a^i_{3PN}[v^2] &=& 
v^2 \left (
-4V^i\dot{U}
-4U^{,j}\Phi_1^{ij}     
-\tfrac{1}{4}\ddot{X}_{1}^{,i}
-\tfrac{1}{2}\ddot{X}_{2}^{,i}
+\tfrac{1}{24}\stackrel{(4)\,\,}{Y^{,i}} 
-\tfrac{1}{8}\Sigma^{,i}(v^4)
-\tfrac{3}{2}\Sigma^{,i}(Uv^2)
+\tfrac{3}{2}\Sigma^{,i}(U^2)
\right .\nonumber \\
&&
\left .
-\tfrac{3}{2}\Sigma^{,i}(\Phi_1)
-\Sigma^{,i}(\Phi_2)
+4\Sigma^{,i}(v^jV^j)
-\tfrac{1}{2}\Sigma^{,i}(\ddot{X})       
-4U^{,j}P_2^{ij}
-4G_{2}^{,i}
+8G_{3}^{,i}
-4G_{6}^{,i}
-4H^{,i}
\right ) ,
\label{eq:v2}
\\
a^i_{3PN}[v^{jk}] &=&  
v^jv^k \bigg( 
4U^{,i}\Phi^{jk}_1       
+16V^jV^{[i,k]}
-8V^iV^{k,j}
+\ddot{X}^{,i}(v^kv^j)
-2\ddot{X}^{,j}(v^iv^k)
-8\Sigma^{,i}(V^kv^j)         
\nonumber \\
&&
+8\Sigma^{,j}(V^kv^i) 
+8\Sigma^{,j}(V^iv^k)    
+\Sigma^{,i}(v^jv^kv^2)     
-2\Sigma^{,j}(v^iv^kv^2)
+6\Sigma^{,i}(Uv^jv^k)
-12\Sigma^{,j}(Uv^iv^k)       
\nonumber \\
&&
\fbox{$+4U^{,i}P_2^{jk}$}
+16P^{,i}(U^{,j}\dot{V}^k)
-32P^{,j}(U^{(,i}\dot{V}^{k)})
+6P^{,i}(\Phi_{1}^{,(j}U^{,k)})
-4P^{,i}(\Phi_{2}^{,(j}U^{,k)})
\nonumber \\
&&
-12P^{,j}(\Phi_{1}^{,(k}U^{,i)})
+8P^{,j}(\Phi_{2}^{,(k}U^{,i)})
-8P^{,i}(V^{m,j}V^{m,k})
+16P^{,j}(V^{m,i}V^{m,k})       
+16P^{,i}(V^{m,j}V^{k,m})
\nonumber \\
&&
-32P^{,j}(V^{m,(i}V^{k),m})
+2P^{,i}(U^{,j}\ddot{X}^{,k})
-4P^{,j}(U^{(,i}\ddot{X}^{,k)})
+\ddot{S}^{,i}(U^{,j}U^{,k})
-2\ddot{S}^{,j}(U^{,i}U^{,k})
\bigg) \,,
\label{eq:vjvk}
\\
%
a^i_{3PN}[v^i] &=&
v^i \bigg( 
8U^{,j}UV^j
-8U^{,j}\Phi^j_{2}
+4U^{,j}V^j_{2}
+2U^{,j}V^j_3
+2U^{,j}\ddot{X}^j
+6V^j\Phi_{1}^{,j}
-4V^j\Phi_{2}^{,j}
+2V^j\ddot{X}^{,j}       
\nonumber\\ 
&&
+8\dot{V}^jV^j
-\tfrac{1}{4}\dddot{X}_{1}
+\tfrac{3}{2}\dddot{X}_{2}
-\tfrac{1}{8}\stackrel{(5)\,}{Y}
-\tfrac{5}{8}\dot{\Sigma}(v^4)
-\tfrac{3}{2}\dot{\Sigma}(v^2U)
-\tfrac{9}{2}\dot{\Sigma}(U^2)
+\tfrac{9}{2}\dot{\Sigma}(\Phi_1)
+3\dot{\Sigma}(\Phi_2)
\nonumber\\
&&
-4\dot{\Sigma}(v^jV^j)
+\tfrac{3}{2}\dot{\Sigma}(\ddot{X})
+6\dot{G}_{1}
+12\dot{G}_{2}
-24\dot{G}_{3}
-8\dot{G}_{4}
+12\dot{G}_{6}
+16U^{,j}G_7^j
+12\dot{H}
\bigg) \,,
\label{eq:vi}
\\
%
%
a^i_{3PN}[v^j] &= & v^j \bigg(
16UU^{,i}V^j
+32U^2V^{[i,j]}
-32U\Phi_2^{[j,i]}
+16UV_2^{[j,i]}       
+8UV_3^{[j,i]}  
-16U^{,j}\Phi_2^i
+8U^{,j}V_2^i
+4U^{,j}V_{3}^i   
\nonumber \\
&&
-8V^{[i,j]}\Phi_1
+16V^{[i,j]}\Phi_2 
+8V^i\dot{V}^j
+8V^j\dot{V}^i  
+4V^i\Phi_{1,j}
+8V^j\Phi_{1,i}
-8V^i\Phi_{2,j} 
+16V^k\Phi_{1}^{k[i,j]}
\nonumber \\
&&
-8V^{k,(i}\Phi_1^{j)k}    
+16V^{j,k}\Phi_1^{ik}   
-8U\ddot{X}^{[i,j]}  
+4U^{,j}\ddot{X}^i
+4V^i\ddot{X}^{,j}
-8V^{[i,j]}\ddot{X}
+4\Sigma^{,[i}(v^{j]}U^2)
\nonumber \\
&&
-20\Sigma^{,[i}(v^{j]}v^2U)
+3\Sigma^{,[j}(v^{i]}v^4)  
+4\Sigma^{,[i}(v^{j]}\Phi_1)      
+24\Sigma^{,[i}(v^{j]}\Phi_2)
-16\Sigma^{,[i}(v^k\Phi_1^{j]k}) 
+8\Sigma^{,[i}(v^2V^{j]})
\nonumber \\
&&    
+16\Sigma^{,[i}(v^{j]}v^kV^k)      
+8\Sigma^{,[i}(UV^{j]})        
-32\Sigma^{,[i}(\Phi_2^{j]})  
+16\Sigma^{,[i}(V_2^{j]})
+8\Sigma^{,[i}(V_3^{j]})  
+8\Sigma^{,[i}(\ddot{X}(v^{j]})) 
\nonumber \\
&&
-4\Sigma^{,[i}(v^{j]}\ddot{X})       
+16\dot{\Sigma}(v^{(i}V^{j)}) 
-2\dot{\Sigma}(v^iv^jv^2)
-12\dot{\Sigma}(v^iv^jU)
-2\ddot{X}^{,[i}(v^{j]}v^2)       
-2\ddot{X}^{,[i}(v^{j]}U)       
\nonumber \\
&&
+8\ddot{X}^{,[i}(V^{j]})     
-2 \dddot{X}(v^iv^j)
-\tfrac{1}{3}\stackrel{(4)\,\,\,}{Y^{,[i}}(v^{j]})     
-64UG_{7}^{[i,j]}
+32U_{,j}G^i_7
+16V^kP_2^{k[i,j]}
-16V^{k,(i}P_2^{j)k}
\nonumber\\
&&
+16V^j_{,k}P_2^{ik}
-4P^{,[i}(\dot{U}\Phi_{1}^{,j]})   
-4P^{,[i}(U^{,j]}\dot{\Phi}_{1})
+24P^{,[i}(\dot{U}\Phi_{2}^{,j]})
+24P^{,[i}(U^{,j]}\dot{\Phi}_{2})
-32P^{,[i}(U^{,k}\dot{\Phi}_{1}^{j]k})         
\nonumber\\
&&
+64P^{[,i}(U^{,k}\Phi_2^{k,j]}) 
-32P^{,[i}(U^{,k}V_2^{k,j]})   
-16P^{,[i}(U^{,k}V_3^{k,j]})      
-16P^{,[i}(V^{k,j]}\Phi_{1}^{,k})   
-32P^{,[i}(V^{k,m}\Phi_{1}^{j]m,k})
\nonumber\\
&&
+32P^{,[i}\Phi_{1}^{km,j]} (V^{k,m})  
+32P^{,[i}(V^{k,j]}\Phi_{2}^{,k})
+16\dot{P}(V^{k,i}V^{k,j})       
-32\dot{P}(V^{k,(i}V^{j),k})
-12\dot{P}(U^{,(i}\Phi_{1}^{,j)})
\nonumber\\
&&
+8\dot{P}(U^{,(i}\Phi_{2}^{,j)})
-32\dot{P}(U_{,(i}V_{j),t})
+64P^{,[i}( \dot{V}^{j],k} V^k)        
+32P^{,[i}(\ddot{V}^{j]}U)      
+32P^{,[i}(V^{j],lm}\Phi_1^{lm})  
\nonumber\\
&&
+64\Sigma^{,[i}(G_7^{j]})            
-16\Sigma^{,[i}(v^kP_2^{j]k})
-12P^{,[i}(U^{,j]} \dddot{X})      
-12P^{,[i}( \ddot{X}^{,j]} \dot{U})            
-16P^{,[i}( \ddot{X}^{,j]}(v^k) U^{,k})    
\nonumber\\
&&
-16P^{,[i}(V^{k,j]} \ddot{X}^{,k})   
-4\dot{P}(U^{,(i} \ddot{X}^{,j)})
-2 \dddot{S}(U^{,i}U^{,j})    
-16 \ddot{S}^{,[i}(V^{k,j]} U^{,k})      
-12 \ddot{S}^{,[i}(U^{,j]} \dot{U})       
\nonumber\\
&&
+32P^{,[i}(P_{2}^{km,j]} V^{k,m})
-32P^{,[i}(P_{2}^{j]m,k} V^{k,m})
-32P^{,[i}( \dot{P}_{2}^{j]k} U^{,k})      
-128P^{,[i}(G_7^{k,j]} U^{,k}) 
\nonumber\\
&&
+32P^{,[i}(V^{j],lm}P_2^{lm})
-64P^{,[i}(V^{j],k}UU^{,k})
-48P^{,[i}(U^{,j]}U^{,k}V^k)      
-48P^{,[i}(U^{,j]}U \dot{U})
\bigg) \,,
\label{eq:vj}
\\
%
a^i_{3PN}(v^0)&=&
 \fbox{- $\tfrac{32}{3}U^3U^{,i}$}
+8UU^{,i}\Phi_1
+12U^2\Phi_{1}^{,i}         
-16UU^{,i}\Phi_2
-8U^2\Phi_{2}^{,i}
+16UU^{,j}\Phi^{ij}_{1}  
-8U^{,j}V^iV^j            
-16UV^jV^{j,i}   
\nonumber \\
&&
+16U^2 \dot{V}^i
+8U \dot{U}V^i
+4U^2 \ddot{X}^{,i}
+8UU^{,i} \ddot{X}
-\tfrac{7}{2}U\Sigma^{,i}(v^4)
-18U\Sigma^{,i}(v^2U)
-6U\Sigma^{,i}(U^2)        
+6U\Sigma^{,i}(\Phi_1)  
\nonumber \\
&& 
+4U\Sigma^{,i}(\Phi_2)      
+16U\Sigma^{,i}(v^jV^j)     
-\tfrac{3}{2}U^{,i}\Sigma(v^4)
-6U^{,i}\Sigma(v^2U)
-6U^{,i}\Sigma(U^2)
+6U^{,i}\Sigma(\Phi_1)
+4U^{,i}\Sigma(\Phi_2)
\nonumber \\
&&
+16U^{,j}\Sigma(v^{(i}V^{j)})
-2U^{,j}\Sigma(v^iv^jv^2)
-12U^{,j}\Sigma(v^iv^jU)
+16U \dot{\Phi}^i_{2}         
-8U \dot{V}^i_{2}
-4U \dot{V}^i_{3} 
-8 \dot{U}\Phi_2^i
\nonumber \\
&&              
+4\dot{U}V_2^i
+2\dot{U}V_3^i
-3\Phi_{1}^{,i}\Phi_1
+2\Phi_{2}^{,i}\Phi_1
+6\Phi_{1}^{,i}\Phi_2       
-4\Phi_{2}^{,i}\Phi_2
-6\Phi_{1}^{,j}\Phi_1^{ij}
+4\Phi_{2}^{,j}\Phi_1^{ij}     
-4 \dot{V}^i  \Phi_1          
+8 \dot{V}^i \Phi_2
\nonumber \\
&&
-8 \dot{V}^j \Phi_1^{ij}
-2V^i \dot{\Phi}_{1}             
-4V^i \dot{\Phi}_{2}
+8V^j \dot{\Phi}_{1}^{ij}     
-16V^j \Phi_2^{j,i}
+8V^j V_2^{j,i}
+4V^jV_3^{j,i}            
-16V^{j,i}\Phi^j_2      
+8V^{j,i}V_2^j
\nonumber \\
&&
+4V^{j,i}V^j_3             
-4U \dddot{X}^i
-3U \ddot{X}_{1}^{,i}
+2U\ddot{X}_{2}^{,i}           
+2 \dot{U} \ddot{X}^i
-U^{,i} \ddot{X}_{1}
+2U^{,i} \ddot{X}_{2}
-2U^{,j} \ddot{X}(v^iv^j)
-\Phi_1  \ddot{X}^{,i}
\nonumber \\
&&
+2\Phi_2 \ddot{X}^{,i}
-2\Phi_1^{ij} \ddot{X}^{,j}
-3\Phi_{1}^{,i} \ddot{X}
+2\Phi_{2}^{,i} \ddot{X}
+2U^{,i}\Sigma( \ddot{X})
+2U\Sigma^{,i}(\ddot{X}) 
+2V^i \dddot{X}
-4 \dot{V}^i \ddot{X}
+4V^j \ddot{X}^{j,i}
\nonumber \\
&&
+4V^{j,i} \ddot{X}^j
- \ddot{X}^{,i}\ddot{X}        
-\tfrac{1}{6}U \stackrel{(4)\,\,}{Y^{,i}}
-\tfrac{1}{6}U^{,i} \stackrel{(4)}{Y}
+\tfrac{11}{16}\Sigma^{,i}(v^6)
+\tfrac{49}{8}\Sigma^{,i}(v^4U) 
+\tfrac{33}{4}\Sigma^{,i}(v^2U^2)
-\tfrac{7}{6}\Sigma^{,i}(U^3)
\nonumber \\
&&
+\tfrac{3}{4}\Sigma^{,i}(v^2\Phi_1)
-\tfrac{15}{2}\Sigma^{,i}(v^2\Phi_2)
+6\Sigma^{,i}(v^jv^k\Phi_1^{jk})   
+\tfrac{9}{2}\Sigma^{,i}(U\Phi_1)
-\Sigma^{,i}(U\Phi_2)
-10\Sigma^{,i}(v^jV^jv^2)
\nonumber \\
&&      
-20\Sigma^{,i}(v^jV^jU)         
+8\Sigma^{,i}(v^j\Phi_2^j)
-4\Sigma^{,i}(v^jV_2^j)
-2\Sigma^{,i}(v^jV_3^j)
-\tfrac{7}{8}\Sigma^{,i}(\Sigma(v^4))
-\tfrac{15}{2}\Sigma^{,i}(\Sigma(v^2U))
\nonumber \\
&&
+\tfrac{1}{2}\Sigma^{,i}(\Sigma(U^2))
-\tfrac{3}{2}\Sigma^{,i}(\Sigma(\Phi_1))   
+3\Sigma^{,i}(\Sigma(\Phi_2))
+12\Sigma^{,i}(\Sigma(v^jV^j)) 
+\tfrac{3}{2} \dot{\Sigma}(v^iv^4)
-2\dot{\Sigma}(v^iU^2)
+10\dot{\Sigma}(v^iv^2U)  
\nonumber \\
&&          
-2\dot{\Sigma}(v^i\Phi_1)
-12 \dot{\Sigma}(v^i\Phi_2)     
+8\dot{\Sigma}(v^j\Phi_1^{ij})
-8\dot{\Sigma}(UV^i)              
-4\dot{\Sigma}(v^2V^i)
-8\dot{\Sigma}(v^iv^kV^k)
+16\dot{\Sigma}(\Phi_2^i)  
\nonumber \\
&&             
-8\dot{\Sigma}(V_2^i)
-4\dot{\Sigma}(V_3^i)
+\tfrac{7}{16}\ddot{X}^{,i}(v^4)  
+\tfrac{9}{4}\ddot{X}^{,i}(v^2U)         
+\tfrac{3}{4}\ddot{X}^{,i}(U^2)
-\tfrac{3}{4}\ddot{X}^{,i}(\Phi_1)
-\tfrac{1}{2}\ddot{X}^{,i}(\Phi_2)
-2\ddot{X}^{,i}(v^jV^j)
\nonumber \\
&&
+\dddot{X}(v^iv^2)         
+2\dddot{X}(v^iU)                    
-4\dddot{X}(V^i)            
-\tfrac{1}{4}\ddot{X}^{,i}(\ddot{X})
+\tfrac{3}{2}\Sigma^{,i}(U\ddot{X})   
+\tfrac{9}{4}\Sigma^{,i}(v^2\ddot{X})
-\tfrac{3}{4}\Sigma^{,i}(\ddot{X}_{1}) 
-\tfrac{1}{2}\Sigma^{,i}(\ddot{X}_{2}) 
\nonumber \\
&& 
-2\Sigma^{,i}(v^j\ddot{X}^j)
-\tfrac{1}{2}\Sigma^{,i}(\Sigma(\ddot{X}))    
+2\dot{\Sigma}(v^i\ddot{X})
-4\dot{\Sigma}(\ddot{X}(v^i))
-\tfrac{1}{24} \stackrel{(4)\,\,}{Y^{,i}}(U)          
+\tfrac{1}{16}\stackrel{(4)\,\,}{Y^{,i}}(v^2)
+\tfrac{1}{6}\stackrel{(5)}{Y}(v^i)
\nonumber \\
&&
-\tfrac{1}{24}\Sigma^{,i}(\stackrel{(4)}{Y})
+\tfrac{1}{720} \stackrel{(6)\,\,}{Z^{,i}}
+16UU^{,j}P_2^{ij}         
+12U^{,i}G_1
+16U^{,i}G_2
-32U^{,i}G_3
-16U^{,i}G_4
+16U^{,i}G_6
\nonumber \\
&&
+24UG_{1}^{,i}
+16UG_{2}^{,i}
-32UG_{3}^{,i}
-32UG_{4}^{,i}
+16UG_{6}^{,i}
-32U \dot{G}^i_{7}
+16 \dot{U}G_7^i
+32V^jG_{7}^{j,i}
+32V^{j,i}G^j_7
\nonumber \\
&&
-12U^{,j}P(\Phi_{1}^{,(i}U^{,j)})
+8U^{,j}P(\Phi_{2}^{,(i}U^{,j)})
+16U^{,j}P(V^{k,i}V^{k,j})
-32U^{,j}P(V^{k,(i}V^{j),k})
-32U^{,j}P(U^{,(i} \dot{V}^{j)})
\nonumber \\
&&
+8V^j \dot{P}_{2}^{ij}
-8 \dot{V}^j P_2^{ij}                
-6\Phi_{1}^{,j}P_2^{ij}
+4\Phi_{2}^{,j}P_2^{ij}       
-2 \ddot{X}^{,j}P_2^{ij}
-2P^{,i}( \dot{U} \dot{\Phi}_{1})          
+12P^{,i}( \dot{U} \dot{\Phi}_{2})  
+8P^{,i}( \dot{V}^j \dot{V}^j)             
\nonumber \\
&&
+16P^{,i}(V^{j,k} \dot{\Phi}^{jk}_{1})  
-32P^{,i}(V^{j,k}\Phi_{2}^{k,j})   
+16P^{,i}(V^{j,k}V_{2}^{k,j})
+8P^{,i}(V^{j,k}V_{3}^{k,j}) 
+8\dot{P}(V^{j,i}\Phi_{1}^{,j})
\nonumber \\
&&
+16\dot{P}(V^{j,k}\Phi_{1}^{ik,j})          
-16\dot{P}(V^{k,j}\Phi_{1}^{kj,i})         
-16\dot{P}(V^{j,i}\Phi_{2}^{,j})   
+2\dot{P}(\dot{U} \Phi_{1}^{,i})
+2\dot{P}(U^{,i} \dot{\Phi}_{1})          
+16\dot{P}(U^{,j} \dot{\Phi}_{1}^{ij})        
\nonumber \\
&&
-12\dot{P}(\dot{U}\Phi_{2}^{,i})    
-12\dot{P}(U^{,i} \dot{\Phi}_{2})
-32\dot{P}(U^{,k}\Phi^{k,i}_{2})
+16\dot{P}(U^{,k}V^{k,i}_{2})        
+8\dot{P}(U^{,k}V^{k,i}_{3})       
-2P^{,i}(\ddot{U}\Phi_1)
\nonumber \\
&&
+4P^{,i}(\ddot{U}\Phi_2)
-6P^{,i}(U\ddot{\Phi}_{1})
+4P^{,i}(U \ddot{\Phi}_{2})
-12P^{,i}(\dot{\Phi}_{1}^{,j}V^j)
+8P^{,i}(\dot{\Phi}_{2}^{,j}V^j)
+16P^{,i}(\dot{U}^{,j}\Phi_2^j)
\nonumber \\
&&
-8P^{,i}(\dot{U}^{,j}V_2^j)          
-4P^{,i}(\dot{U}^{,j}V^j_3)
-12P^{,i}(U^{,jk}\Sigma(v^jv^kU))
-2P^{,i}(U^{,jk}\Sigma(v^jv^kv^2))
-6P^{,i}(\Phi_{1}^{,jk}\Phi^{jk}_1)
\nonumber \\
&&
+4P^{,i}(\Phi_{2}^{,jk}\Phi^{jk}_1)
-16\dot{P}(\ddot{V}^iU)
-32\dot{P}(V^k \dot{V}^{i,k})
-16\dot{P}(V^{i,jl}\Phi_1^{jl})  
+6\Sigma^{,i}(v^jv^kP_2^{jk})       
+8\dot{\Sigma}(v^jP_2^{ij})  
\nonumber \\
&&
-2\Sigma^{,i}(G_1)
+4\Sigma^{,i}(G_2)
-8\Sigma^{,i}(G_3)          
-16\Sigma^{,i}(G_4)
+16\Sigma^{,i}(G_5)
+4\Sigma^{,i}(G_6)
-16\Sigma^{,i}(v^jG^j_7)    
-32\dot{\Sigma}(G^i_7)
\nonumber \\
&&
-4 U^{,j}P(U^{,(i} \ddot{X}^{,j)})
+8\dot{P}(U^{,k}\ddot{X}^{,i}(v^k))
+6\dot{P}(U^{,i}\dddot{X})
-6P^{,i}(\dot{U} \dddot{X})
+8P^{,i}(V^{k,j}\ddot{X}^{j,k})
+8\dot{P}(V^{k,i}\ddot{X}^{,k})
\nonumber \\
&&
+6\dot{P}(\dot{U}\ddot{X}^{,i})
-4P^{,i}(\dot{U}^{,j} \ddot{X}^j)         
-2P^{,i}(U \stackrel{(4)}{X})          
-4P^{,i}(V^j \dddot{X}^{,j})
-2P^{,i}(\ddot{U}\ddot{X})         
-2P^{,i}( \ddot{X}^{,jk}\Phi^{jk}_1)
\nonumber \\
&&
-2P^{,i}(U^{,jk}\ddot{X}^{jk}_{1})
-2 \ddot{S}(U^{,i}U^{,j})U^{,j}
-3 \ddot{S}^{,i}(\dot{U}^2)
+4\ddot{S}^{,i}(V^{j,k}V^{k,j})        
+6\dddot{S}(\dot{U}U^{,i}) 
+8 \dddot{S}(U^{,k}V^{k,i})      
\nonumber \\
&&  
-4\ddot{S}^{,i}(\dot{U}^{,j}V^j)         
-2\ddot{S}^{,i}(\ddot{U}U)
-2\ddot{S}^{,i}(U^{,jk}\Phi^{jk}_1)
+\fbox{$16UH^{,i}$}
+16HU^{,i}
+4\Sigma^{,i}(H)
+16\dot{P}(U^{,j} \dot{P}_{2}^{ij})
\nonumber \\
&&
+64\dot{P}(U^{,k}G_{7}^{k,i})
+16P^{,i}(V^{j,k} \dot{P}^{jk}_{2})    
+32\dot{P}(V^{j,k}P_{2}^{k[i,j]})
+64P^{,i}(V^{k,j}G_7^{j,k})   
-12P^{,i}(U^{,jk}P(U^{,j}\Phi_{1}^{,k}))        
\nonumber \\
&&
+8P^{,i}(U^{,jk}P(U^{,j}\Phi_{2}^{,k}))
-32P^{,i}(U^{,jk}P(U^{,j}\dot{V}^k))
-4P^{,i}(U^{,jk}P(U^{,j} \ddot{X}^{,k}))
-32P^{,i}(\dot{U}^{,j}G_7^j)               
\nonumber \\
&&
+64P^{,i}(U^{,jk}P(V^{[m,j]}V^{[m,k]}))    
-6P^{,i}(\Phi_{1}^{,jk}P_2^{jk})
+4P^{,i}(\Phi_{2}^{,jk} P_2^{jk})
-16\dot{P}(V^{i,jk}P^{jk}_2)  
-24P^{,i}(\dot{U}^2U)      
\nonumber \\
&&
+16P^{,i}(UV^{j,k}V^{j,k})          
+32\dot{P}(U^{,k}V^{i,k}U)        
+24\dot{P}(\dot{U}U^{,i}U)  
+24\dot{P}(U^{,k}U^{,i}V^k)
-24P^{,i}(U^{,k}\dot{U}V^k)                  
\nonumber \\
&&
-16P^{,i}(\dot{U}^{,k}UV^k)
-8P^{,i}(\ddot{U}U^2)
-2P^{,i}(\ddot{X}^{,jk}P_2^{jk})
-2P^{,i}(U^{,jk}\ddot{S}(U^{,j}U^{,k}))
-2\ddot{S}^{,i}(U^{,jk}P_2^{jk})  \,. 
\label{eq:nov}     
\end{eqnarray}
There will be additional 2PN and 3PN terms, generated by inserting the equations of motion into places where time derivatives generate accelerations.  Thus, for example, in the 1PN terms $\dot{V}^i$ and $\ddot{X}^{,j}$, inserting the Newtonian, 1PN and 2PN pieces of the acceleration will generate 1PN, 2PN and 3PN terms.  We have not displayed these terms explicitly.   In addition, the effects of relativistic ``flattening'' of the shape of each body will have to be incorporated.
\end{widetext}


\begin{thebibliography}{38}
\expandafter\ifx\csname natexlab\endcsname\relax\def\natexlab#1{#1}\fi
\expandafter\ifx\csname bibnamefont\endcsname\relax
  \def\bibnamefont#1{#1}\fi
\expandafter\ifx\csname bibfnamefont\endcsname\relax
  \def\bibfnamefont#1{#1}\fi
\expandafter\ifx\csname citenamefont\endcsname\relax
  \def\citenamefont#1{#1}\fi
\expandafter\ifx\csname url\endcsname\relax
  \def\url#1{\texttt{#1}}\fi
\expandafter\ifx\csname urlprefix\endcsname\relax\def\urlprefix{URL }\fi
\providecommand{\bibinfo}[2]{#2}
\providecommand{\eprint}[2][]{\url{#2}}

\bibitem[{\citenamefont{{Abbott} et~al.}(2023)\citenamefont{{Abbott}, {Abbott},
  {Acernese}, {Ackley}, {Adams}, {Adhikari}, {Adhikari}, {Adya}, {Affeldt},
  {Agarwal} et~al.}}]{2023PhRvX..13d1039A}
\bibinfo{author}{\bibfnamefont{R.}~\bibnamefont{{Abbott}}},
  \bibinfo{author}{\bibfnamefont{T.~D.} \bibnamefont{{Abbott}}},
  \bibinfo{author}{\bibfnamefont{F.}~\bibnamefont{{Acernese}}},
  \bibinfo{author}{\bibfnamefont{K.}~\bibnamefont{{Ackley}}},
  \bibinfo{author}{\bibfnamefont{C.}~\bibnamefont{{Adams}}},
  \bibinfo{author}{\bibfnamefont{N.}~\bibnamefont{{Adhikari}}},
  \bibinfo{author}{\bibfnamefont{R.~X.} \bibnamefont{{Adhikari}}},
  \bibinfo{author}{\bibfnamefont{V.~B.} \bibnamefont{{Adya}}},
  \bibinfo{author}{\bibfnamefont{C.}~\bibnamefont{{Affeldt}}},
  \bibinfo{author}{\bibfnamefont{D.}~\bibnamefont{{Agarwal}}},
  \bibnamefont{et~al.}, \bibinfo{journal}{Physical Review X}
  \textbf{\bibinfo{volume}{13}}, \bibinfo{eid}{041039} (\bibinfo{year}{2023}),
  \eprint{2111.03606}.

\bibitem[{\citenamefont{{Agazie} et~al.}(2023)\citenamefont{{Agazie},
  {Anumarlapudi}, {Archibald}, {Arzoumanian}, {Baker}, {B{\'e}csy}, {Blecha},
  {Brazier}, {Brook}, {Burke-Spolaor} et~al.}}]{2023ApJ...951L...8A}
\bibinfo{author}{\bibfnamefont{G.}~\bibnamefont{{Agazie}}},
  \bibinfo{author}{\bibfnamefont{A.}~\bibnamefont{{Anumarlapudi}}},
  \bibinfo{author}{\bibfnamefont{A.~M.} \bibnamefont{{Archibald}}},
  \bibinfo{author}{\bibfnamefont{Z.}~\bibnamefont{{Arzoumanian}}},
  \bibinfo{author}{\bibfnamefont{P.~T.} \bibnamefont{{Baker}}},
  \bibinfo{author}{\bibfnamefont{B.}~\bibnamefont{{B{\'e}csy}}},
  \bibinfo{author}{\bibfnamefont{L.}~\bibnamefont{{Blecha}}},
  \bibinfo{author}{\bibfnamefont{A.}~\bibnamefont{{Brazier}}},
  \bibinfo{author}{\bibfnamefont{P.~R.} \bibnamefont{{Brook}}},
  \bibinfo{author}{\bibfnamefont{S.}~\bibnamefont{{Burke-Spolaor}}},
  \bibnamefont{et~al.}, \bibinfo{journal}{\apjl}
  \textbf{\bibinfo{volume}{951}}, \bibinfo{eid}{L8} (\bibinfo{year}{2023}),
  \eprint{2306.16213}.

\bibitem[{\citenamefont{{Evans} et~al.}(2023)\citenamefont{{Evans}, {Corsi},
  {Afle}, {Ananyeva}, {Arun}, {Ballmer}, {Bandopadhyay}, {Barsotti},
  {Baryakhtar}, {Berger} et~al.}}]{2023arXiv230613745E}
\bibinfo{author}{\bibfnamefont{M.}~\bibnamefont{{Evans}}},
  \bibinfo{author}{\bibfnamefont{A.}~\bibnamefont{{Corsi}}},
  \bibinfo{author}{\bibfnamefont{C.}~\bibnamefont{{Afle}}},
  \bibinfo{author}{\bibfnamefont{A.}~\bibnamefont{{Ananyeva}}},
  \bibinfo{author}{\bibfnamefont{K.~G.} \bibnamefont{{Arun}}},
  \bibinfo{author}{\bibfnamefont{S.}~\bibnamefont{{Ballmer}}},
  \bibinfo{author}{\bibfnamefont{A.}~\bibnamefont{{Bandopadhyay}}},
  \bibinfo{author}{\bibfnamefont{L.}~\bibnamefont{{Barsotti}}},
  \bibinfo{author}{\bibfnamefont{M.}~\bibnamefont{{Baryakhtar}}},
  \bibinfo{author}{\bibfnamefont{E.}~\bibnamefont{{Berger}}},
  \bibnamefont{et~al.}, \bibinfo{journal}{arXiv e-prints}
  \bibinfo{eid}{arXiv:2306.13745} (\bibinfo{year}{2023}), \eprint{2306.13745}.

\bibitem[{\citenamefont{Maggiore et~al.}(2020)\citenamefont{Maggiore, Broeck,
  Bartolo, Belgacem, Bertacca, Bizouard, Branchesi, Clesse, Foffa,
  Garc{\'\i}a-Bellido et~al.}}]{Maggiore_2020}
\bibinfo{author}{\bibfnamefont{M.}~\bibnamefont{Maggiore}},
  \bibinfo{author}{\bibfnamefont{C.~V.~D.} \bibnamefont{Broeck}},
  \bibinfo{author}{\bibfnamefont{N.}~\bibnamefont{Bartolo}},
  \bibinfo{author}{\bibfnamefont{E.}~\bibnamefont{Belgacem}},
  \bibinfo{author}{\bibfnamefont{D.}~\bibnamefont{Bertacca}},
  \bibinfo{author}{\bibfnamefont{M.~A.} \bibnamefont{Bizouard}},
  \bibinfo{author}{\bibfnamefont{M.}~\bibnamefont{Branchesi}},
  \bibinfo{author}{\bibfnamefont{S.}~\bibnamefont{Clesse}},
  \bibinfo{author}{\bibfnamefont{S.}~\bibnamefont{Foffa}},
  \bibinfo{author}{\bibfnamefont{J.}~\bibnamefont{Garc{\'\i}a-Bellido}},
  \bibnamefont{et~al.}, \bibinfo{journal}{Journal of Cosmology and
  Astroparticle Physics} \textbf{\bibinfo{volume}{2020}}, \bibinfo{pages}{050}
  (\bibinfo{year}{2020}), ISSN \bibinfo{issn}{1475-7516},
  \urlprefix\url{http://dx.doi.org/10.1088/1475-7516/2020/03/050}.

\bibitem[{\citenamefont{{Blanchet}}(2024)}]{2024LRR....27....4B}
\bibinfo{author}{\bibfnamefont{L.}~\bibnamefont{{Blanchet}}},
  \bibinfo{journal}{Living Reviews in Relativity}
  \textbf{\bibinfo{volume}{27}}, \bibinfo{eid}{4} (\bibinfo{year}{2024}).

\bibitem[{\citenamefont{{Chatziioannou}}(2020)}]{2020GReGr..52..109C}
\bibinfo{author}{\bibfnamefont{K.}~\bibnamefont{{Chatziioannou}}},
  \bibinfo{journal}{General Relativity and Gravitation}
  \textbf{\bibinfo{volume}{52}}, \bibinfo{eid}{109} (\bibinfo{year}{2020}),
  \eprint{2006.03168}.

\bibitem[{\citenamefont{{Abbott} et~al.}(2017)\citenamefont{{Abbott}, {Abbott},
  {Abbott}, {Acernese}, {Ackley}, {Adams}, {Adams}, {Addesso}, {Adhikari},
  {Adya} et~al.}}]{2017PhRvL.119p1101A}
\bibinfo{author}{\bibfnamefont{B.~P.} \bibnamefont{{Abbott}}},
  \bibinfo{author}{\bibfnamefont{R.}~\bibnamefont{{Abbott}}},
  \bibinfo{author}{\bibfnamefont{T.~D.} \bibnamefont{{Abbott}}},
  \bibinfo{author}{\bibfnamefont{F.}~\bibnamefont{{Acernese}}},
  \bibinfo{author}{\bibfnamefont{K.}~\bibnamefont{{Ackley}}},
  \bibinfo{author}{\bibfnamefont{C.}~\bibnamefont{{Adams}}},
  \bibinfo{author}{\bibfnamefont{T.}~\bibnamefont{{Adams}}},
  \bibinfo{author}{\bibfnamefont{P.}~\bibnamefont{{Addesso}}},
  \bibinfo{author}{\bibfnamefont{R.~X.} \bibnamefont{{Adhikari}}},
  \bibinfo{author}{\bibfnamefont{V.~B.} \bibnamefont{{Adya}}},
  \bibnamefont{et~al.}, \bibinfo{journal}{\prl} \textbf{\bibinfo{volume}{119}},
  \bibinfo{eid}{161101} (\bibinfo{year}{2017}), \eprint{1710.05832}.

\bibitem[{\citenamefont{Damour}(1987)}]{Damour300}
\bibinfo{author}{\bibfnamefont{T.}~\bibnamefont{Damour}}, in
  \emph{\bibinfo{booktitle}{Three Hundred Years of Gravitation}}, edited by
  \bibinfo{editor}{\bibfnamefont{S.~W.} \bibnamefont{Hawking}}
  \bibnamefont{and} \bibinfo{editor}{\bibfnamefont{W.}~\bibnamefont{Israel}}
  (\bibinfo{publisher}{Cambridge University Press},
  \bibinfo{address}{Cambridge}, \bibinfo{year}{1987}), pp.
  \bibinfo{pages}{128--198}.

\bibitem[{\citenamefont{{Will}}(2018)}]{tegp2}
\bibinfo{author}{\bibfnamefont{C.~M.} \bibnamefont{{Will}}},
  \emph{\bibinfo{title}{Theory and Experiment in Gravitational Physics}}
  (\bibinfo{publisher}{Cambridge University Press},
  \bibinfo{address}{Cambridge}, \bibinfo{year}{2018}), \bibinfo{edition}{2nd}
  ed.

\bibitem[{\citenamefont{{Touboul} et~al.}(2017)\citenamefont{{Touboul},
  {M{\'e}tris}, {Rodrigues}, {Andr{\'e}}, {Baghi}, {Berg{\'e}}, {Boulanger},
  {Bremer}, {Carle}, {Chhun} et~al.}}]{PhysRevLett.119.231101}
\bibinfo{author}{\bibfnamefont{P.}~\bibnamefont{{Touboul}}},
  \bibinfo{author}{\bibfnamefont{G.}~\bibnamefont{{M{\'e}tris}}},
  \bibinfo{author}{\bibfnamefont{M.}~\bibnamefont{{Rodrigues}}},
  \bibinfo{author}{\bibfnamefont{Y.}~\bibnamefont{{Andr{\'e}}}},
  \bibinfo{author}{\bibfnamefont{Q.}~\bibnamefont{{Baghi}}},
  \bibinfo{author}{\bibfnamefont{J.}~\bibnamefont{{Berg{\'e}}}},
  \bibinfo{author}{\bibfnamefont{D.}~\bibnamefont{{Boulanger}}},
  \bibinfo{author}{\bibfnamefont{S.}~\bibnamefont{{Bremer}}},
  \bibinfo{author}{\bibfnamefont{P.}~\bibnamefont{{Carle}}},
  \bibinfo{author}{\bibfnamefont{R.}~\bibnamefont{{Chhun}}},
  \bibnamefont{et~al.}, \bibinfo{journal}{\prl} \textbf{\bibinfo{volume}{119}},
  \bibinfo{pages}{231101} (\bibinfo{year}{2017}), \eprint{1712.01176}.

\bibitem[{\citenamefont{Ransom et~al.}(2014)\citenamefont{Ransom, Stairs,
  Archibald, Hessels, Kaplan, van Kerkwijk, Boyles, Deller, Chatterjee,
  Schechtman-Rook et~al.}}]{2014Natur.505..520R}
\bibinfo{author}{\bibfnamefont{S.~M.} \bibnamefont{Ransom}},
  \bibinfo{author}{\bibfnamefont{I.~H.} \bibnamefont{Stairs}},
  \bibinfo{author}{\bibfnamefont{A.~M.} \bibnamefont{Archibald}},
  \bibinfo{author}{\bibfnamefont{J.~W.~T.} \bibnamefont{Hessels}},
  \bibinfo{author}{\bibfnamefont{D.~L.} \bibnamefont{Kaplan}},
  \bibinfo{author}{\bibfnamefont{M.~H.} \bibnamefont{van Kerkwijk}},
  \bibinfo{author}{\bibfnamefont{J.}~\bibnamefont{Boyles}},
  \bibinfo{author}{\bibfnamefont{A.~T.} \bibnamefont{Deller}},
  \bibinfo{author}{\bibfnamefont{S.}~\bibnamefont{Chatterjee}},
  \bibinfo{author}{\bibfnamefont{A.}~\bibnamefont{Schechtman-Rook}},
  \bibnamefont{et~al.}, \bibinfo{journal}{Nature}
  \textbf{\bibinfo{volume}{505}}, \bibinfo{pages}{520} (\bibinfo{year}{2014}),
  \eprint{1401.0535}.

\bibitem[{\citenamefont{{Nordtvedt}}(1968)}]{1968PhRv..169.1017N}
\bibinfo{author}{\bibfnamefont{K.}~\bibnamefont{{Nordtvedt}},
  \bibfnamefont{Jr.}}, \bibinfo{journal}{Phys. Rev.}
  \textbf{\bibinfo{volume}{169}}, \bibinfo{pages}{1017} (\bibinfo{year}{1968}).

\bibitem[{\citenamefont{{Will}}(1971)}]{1971ApJ...163..611W}
\bibinfo{author}{\bibfnamefont{C.~M.} \bibnamefont{{Will}}},
  \bibinfo{journal}{\apj} \textbf{\bibinfo{volume}{163}}, \bibinfo{pages}{611}
  (\bibinfo{year}{1971}).

\bibitem[{\citenamefont{Mitchell and Will}(2007)}]{2007PhRvD..75l4025M}
\bibinfo{author}{\bibfnamefont{T.}~\bibnamefont{Mitchell}} \bibnamefont{and}
  \bibinfo{author}{\bibfnamefont{C.~M.} \bibnamefont{Will}},
  \bibinfo{journal}{Phys. Rev. D} \textbf{\bibinfo{volume}{75}},
  \bibinfo{eid}{124025} (\bibinfo{year}{2007}), \eprint{0704.2243}.

\bibitem[{\citenamefont{Grishchuk and Kopeikin}(1983)}]{1983SvAL....9..230G}
\bibinfo{author}{\bibfnamefont{L.~P.} \bibnamefont{Grishchuk}}
  \bibnamefont{and} \bibinfo{author}{\bibfnamefont{S.~M.}
  \bibnamefont{Kopeikin}}, \bibinfo{journal}{Sov. Astron. Lett.}
  \textbf{\bibinfo{volume}{9}}, \bibinfo{pages}{230} (\bibinfo{year}{1983}).

\bibitem[{\citenamefont{{Will} and {Wiseman}}(1996)}]{1996PhRvD..54.4813W}
\bibinfo{author}{\bibfnamefont{C.~M.} \bibnamefont{{Will}}} \bibnamefont{and}
  \bibinfo{author}{\bibfnamefont{A.~G.} \bibnamefont{{Wiseman}}},
  \bibinfo{journal}{\prd} \textbf{\bibinfo{volume}{54}}, \bibinfo{pages}{4813}
  (\bibinfo{year}{1996}), \eprint{gr-qc/9608012}.

\bibitem[{\citenamefont{{Pati} and {Will}}(2000)}]{2000PhRvD..62l4015P}
\bibinfo{author}{\bibfnamefont{M.~E.} \bibnamefont{{Pati}}} \bibnamefont{and}
  \bibinfo{author}{\bibfnamefont{C.~M.} \bibnamefont{{Will}}},
  \bibinfo{journal}{\prd} \textbf{\bibinfo{volume}{62}}, \bibinfo{eid}{124015}
  (\bibinfo{year}{2000}), \eprint{gr-qc/0007087}.

\bibitem[{\citenamefont{{Pati} and {Will}}(2002)}]{2002PhRvD..65j4008P}
\bibinfo{author}{\bibfnamefont{M.~E.} \bibnamefont{{Pati}}} \bibnamefont{and}
  \bibinfo{author}{\bibfnamefont{C.~M.} \bibnamefont{{Will}}},
  \bibinfo{journal}{\prd} \textbf{\bibinfo{volume}{65}}, \bibinfo{eid}{104008}
  (\bibinfo{year}{2002}), \eprint{gr-qc/0201001}.

\bibitem[{\citenamefont{{Blanchet} and {Damour}}(1986)}]{1986RSPSA.320..379B}
\bibinfo{author}{\bibfnamefont{L.}~\bibnamefont{{Blanchet}}} \bibnamefont{and}
  \bibinfo{author}{\bibfnamefont{T.}~\bibnamefont{{Damour}}},
  \bibinfo{journal}{Proc. R. Soc. A} \textbf{\bibinfo{volume}{320}},
  \bibinfo{pages}{379} (\bibinfo{year}{1986}).

\bibitem[{\citenamefont{{Blanchet}}(1987)}]{1987RSPSA.409..383B}
\bibinfo{author}{\bibfnamefont{L.}~\bibnamefont{{Blanchet}}},
  \bibinfo{journal}{Proceedings of the Royal Society of London Series A}
  \textbf{\bibinfo{volume}{409}}, \bibinfo{pages}{383} (\bibinfo{year}{1987}).

\bibitem[{\citenamefont{Poisson and Will}(2014)}]{PW2014}
\bibinfo{author}{\bibfnamefont{E.}~\bibnamefont{Poisson}} \bibnamefont{and}
  \bibinfo{author}{\bibfnamefont{C.~M.} \bibnamefont{Will}},
  \emph{\bibinfo{title}{Gravity: Newtonian, Post-Newtonian, Relativistic}}
  (\bibinfo{publisher}{Cambridge University Press},
  \bibinfo{address}{Cambridge}, \bibinfo{year}{2014}).

\bibitem[{\citenamefont{{Blanchet}}(2014)}]{2014LRR....17....2B}
\bibinfo{author}{\bibfnamefont{L.}~\bibnamefont{{Blanchet}}},
  \bibinfo{journal}{\lrr} \textbf{\bibinfo{volume}{17}}, \bibinfo{eid}{2}
  (\bibinfo{year}{2014}), \eprint{1310.1528}.

\bibitem[{\citenamefont{{Blanchet}
  et~al.}(1995{\natexlab{a}})\citenamefont{{Blanchet}, {Damour}, {Iyer},
  {Will}, and {Wiseman}}}]{1995PhRvL..74.3515B}
\bibinfo{author}{\bibfnamefont{L.}~\bibnamefont{{Blanchet}}},
  \bibinfo{author}{\bibfnamefont{T.}~\bibnamefont{{Damour}}},
  \bibinfo{author}{\bibfnamefont{B.~R.} \bibnamefont{{Iyer}}},
  \bibinfo{author}{\bibfnamefont{C.~M.} \bibnamefont{{Will}}},
  \bibnamefont{and} \bibinfo{author}{\bibfnamefont{A.~G.}
  \bibnamefont{{Wiseman}}}, \bibinfo{journal}{\prl}
  \textbf{\bibinfo{volume}{74}}, \bibinfo{pages}{3515}
  (\bibinfo{year}{1995}{\natexlab{a}}), \eprint{gr-qc/9501027}.

\bibitem[{\citenamefont{{Blanchet}
  et~al.}(1995{\natexlab{b}})\citenamefont{{Blanchet}, {Damour}, and
  {Iyer}}}]{1995PhRvD..51.5360B}
\bibinfo{author}{\bibfnamefont{L.}~\bibnamefont{{Blanchet}}},
  \bibinfo{author}{\bibfnamefont{T.}~\bibnamefont{{Damour}}}, \bibnamefont{and}
  \bibinfo{author}{\bibfnamefont{B.~R.} \bibnamefont{{Iyer}}},
  \bibinfo{journal}{\prd} \textbf{\bibinfo{volume}{51}}, \bibinfo{pages}{5360}
  (\bibinfo{year}{1995}{\natexlab{b}}), \eprint{gr-qc/9501029}.

\bibitem[{\citenamefont{{Itoh} and {Futamase}}(2003)}]{2003PhRvD..68l1501I}
\bibinfo{author}{\bibfnamefont{Y.}~\bibnamefont{{Itoh}}} \bibnamefont{and}
  \bibinfo{author}{\bibfnamefont{T.}~\bibnamefont{{Futamase}}},
  \bibinfo{journal}{\prd} \textbf{\bibinfo{volume}{68}}, \bibinfo{eid}{121501}
  (\bibinfo{year}{2003}), \eprint{gr-qc/0310028}.

\bibitem[{\citenamefont{{Itoh}}(2004)}]{2004PhRvD..69f4018I}
\bibinfo{author}{\bibfnamefont{Y.}~\bibnamefont{{Itoh}}},
  \bibinfo{journal}{\prd} \textbf{\bibinfo{volume}{69}}, \bibinfo{eid}{064018}
  (\bibinfo{year}{2004}), \eprint{gr-qc/0310029}.

\bibitem[{\citenamefont{{Itoh} et~al.}(2000)\citenamefont{{Itoh}, {Futamase},
  and {Asada}}}]{2000PhRvD..62f4002I}
\bibinfo{author}{\bibfnamefont{Y.}~\bibnamefont{{Itoh}}},
  \bibinfo{author}{\bibfnamefont{T.}~\bibnamefont{{Futamase}}},
  \bibnamefont{and} \bibinfo{author}{\bibfnamefont{H.}~\bibnamefont{{Asada}}},
  \bibinfo{journal}{\prd} \textbf{\bibinfo{volume}{62}}, \bibinfo{eid}{064002}
  (\bibinfo{year}{2000}), \eprint{gr-qc/9910052}.

\bibitem[{\citenamefont{{Einstein} et~al.}(1938)\citenamefont{{Einstein},
  {Infeld}, and {Hoffmann}}}]{1938AnMat..39...65E}
\bibinfo{author}{\bibfnamefont{A.}~\bibnamefont{{Einstein}}},
  \bibinfo{author}{\bibfnamefont{L.}~\bibnamefont{{Infeld}}}, \bibnamefont{and}
  \bibinfo{author}{\bibfnamefont{B.}~\bibnamefont{{Hoffmann}}},
  \bibinfo{journal}{Ann. Math.} \textbf{\bibinfo{volume}{39}},
  \bibinfo{pages}{65} (\bibinfo{year}{1938}).

\bibitem[{\citenamefont{Jaranowski and Sch{\"{a}}fer}(1998)}]{jaraschaefer98}
\bibinfo{author}{\bibfnamefont{P.}~\bibnamefont{Jaranowski}} \bibnamefont{and}
  \bibinfo{author}{\bibfnamefont{G.}~\bibnamefont{Sch{\"{a}}fer}},
  \bibinfo{journal}{Phys. Rev. D} \textbf{\bibinfo{volume}{57}},
  \bibinfo{pages}{7274} (\bibinfo{year}{1998}), \bibinfo{note}{erratum: Phys.
  Rev. D 63 (2001) 029902}, \eprint{gr-qc/9712075}.

\bibitem[{\citenamefont{Jaranowski and Sch{\"{a}}fer}(1999)}]{jaranowski}
\bibinfo{author}{\bibfnamefont{P.}~\bibnamefont{Jaranowski}} \bibnamefont{and}
  \bibinfo{author}{\bibfnamefont{G.}~\bibnamefont{Sch{\"{a}}fer}},
  \bibinfo{journal}{Phys. Rev. D} \textbf{\bibinfo{volume}{60}},
  \bibinfo{eid}{124003} (\bibinfo{year}{1999}), \eprint{gr-qc/9906092}.

\bibitem[{\citenamefont{{Foffa} and {Sturani}}(2011)}]{2011PhRvD..84d4031F}
\bibinfo{author}{\bibfnamefont{S.}~\bibnamefont{{Foffa}}} \bibnamefont{and}
  \bibinfo{author}{\bibfnamefont{R.}~\bibnamefont{{Sturani}}},
  \bibinfo{journal}{\prd} \textbf{\bibinfo{volume}{84}}, \bibinfo{eid}{044031}
  (\bibinfo{year}{2011}), \eprint{1104.1122}.

\bibitem[{\citenamefont{{Blanchet} and {Damour}}(1989)}]{1989AIHPA..50..377B}
\bibinfo{author}{\bibfnamefont{L.}~\bibnamefont{{Blanchet}}} \bibnamefont{and}
  \bibinfo{author}{\bibfnamefont{T.}~\bibnamefont{{Damour}}},
  \bibinfo{journal}{Ann. Inst. Henri Poincar\'e A}
  \textbf{\bibinfo{volume}{50}}, \bibinfo{pages}{377} (\bibinfo{year}{1989}).

\bibitem[{\citenamefont{{Sch{\"a}fer}}(1987)}]{1987PhLA..123..336S}
\bibinfo{author}{\bibfnamefont{G.}~\bibnamefont{{Sch{\"a}fer}}},
  \bibinfo{journal}{Physics Letters A} \textbf{\bibinfo{volume}{123}},
  \bibinfo{pages}{336} (\bibinfo{year}{1987}).

\bibitem[{\citenamefont{{The LIGO Scientific Collaboration}
  et~al.}(2021)\citenamefont{{The LIGO Scientific Collaboration}, {the Virgo
  Collaboration}, {the KAGRA Collaboration}, {Abbott}, {Abe}, {Acernese},
  {Ackley}, {Adhikari}, {Adhikari}, {Adkins} et~al.}}]{2021arXiv211206861T}
\bibinfo{author}{\bibnamefont{{The LIGO Scientific Collaboration}}},
  \bibinfo{author}{\bibnamefont{{the Virgo Collaboration}}},
  \bibinfo{author}{\bibnamefont{{the KAGRA Collaboration}}},
  \bibinfo{author}{\bibfnamefont{R.}~\bibnamefont{{Abbott}}},
  \bibinfo{author}{\bibfnamefont{H.}~\bibnamefont{{Abe}}},
  \bibinfo{author}{\bibfnamefont{F.}~\bibnamefont{{Acernese}}},
  \bibinfo{author}{\bibfnamefont{K.}~\bibnamefont{{Ackley}}},
  \bibinfo{author}{\bibfnamefont{N.}~\bibnamefont{{Adhikari}}},
  \bibinfo{author}{\bibfnamefont{R.~X.} \bibnamefont{{Adhikari}}},
  \bibinfo{author}{\bibfnamefont{V.~K.} \bibnamefont{{Adkins}}},
  \bibnamefont{et~al.}, \bibinfo{journal}{arXiv e-prints}
  \bibinfo{eid}{arXiv:2112.06861} (\bibinfo{year}{2021}), \eprint{2112.06861}.

\bibitem[{\citenamefont{{Perkins} et~al.}(2021)\citenamefont{{Perkins},
  {Yunes}, and {Berti}}}]{2021PhRvD.103d4024P}
\bibinfo{author}{\bibfnamefont{S.~E.} \bibnamefont{{Perkins}}},
  \bibinfo{author}{\bibfnamefont{N.}~\bibnamefont{{Yunes}}}, \bibnamefont{and}
  \bibinfo{author}{\bibfnamefont{E.}~\bibnamefont{{Berti}}},
  \bibinfo{journal}{\prd} \textbf{\bibinfo{volume}{103}}, \bibinfo{eid}{044024}
  (\bibinfo{year}{2021}), \eprint{2010.09010}.

\bibitem[{\citenamefont{{Finstad} et~al.}(2023)\citenamefont{{Finstad},
  {White}, and {Brown}}}]{2023ApJ...955...45F}
\bibinfo{author}{\bibfnamefont{D.}~\bibnamefont{{Finstad}}},
  \bibinfo{author}{\bibfnamefont{L.~V.} \bibnamefont{{White}}},
  \bibnamefont{and} \bibinfo{author}{\bibfnamefont{D.~A.}
  \bibnamefont{{Brown}}}, \bibinfo{journal}{\apj}
  \textbf{\bibinfo{volume}{955}}, \bibinfo{eid}{45} (\bibinfo{year}{2023}),
  \eprint{2211.01396}.

\bibitem[{\citenamefont{{Wang} and {Will}}(2007)}]{2007PhRvD..75f4017W}
\bibinfo{author}{\bibfnamefont{H.}~\bibnamefont{{Wang}}} \bibnamefont{and}
  \bibinfo{author}{\bibfnamefont{C.~M.} \bibnamefont{{Will}}},
  \bibinfo{journal}{\prd} \textbf{\bibinfo{volume}{75}}, \bibinfo{eid}{064017}
  (\bibinfo{year}{2007}), \eprint{gr-qc/0701047}.

\bibitem[{\citenamefont{{Zeng} and {Will}}(2007)}]{2007GReGr..39.1661Z}
\bibinfo{author}{\bibfnamefont{J.}~\bibnamefont{{Zeng}}} \bibnamefont{and}
  \bibinfo{author}{\bibfnamefont{C.~M.} \bibnamefont{{Will}}},
  \bibinfo{journal}{General Relativity and Gravitation}
  \textbf{\bibinfo{volume}{39}}, \bibinfo{pages}{1661} (\bibinfo{year}{2007}),
  \eprint{0704.2720}.

\end{thebibliography}

\end{document}